\documentclass[preprint,12pt]{elsarticle}
\journal{Computational Mechanics}

\usepackage[utf8]{inputenc}
\usepackage[T1]{fontenc}
\usepackage[english]{babel}

\usepackage{microtype}
\usepackage{amsfonts}
\usepackage{amssymb}
\usepackage{amsmath}
\usepackage{graphicx}
\usepackage{caption}
\usepackage{subcaption}
\usepackage{multirow}					
\usepackage{longtable}
\usepackage{booktabs}
\usepackage{float}
\usepackage{pgfplots}
\usepackage{pgfplotstable}
\usepackage{mdframed}
\pgfplotsset{compat=newest}
\usepgfplotslibrary{groupplots}
\usetikzlibrary{spy}
\usepackage[margin=1in]{geometry}
\usepackage[section]{placeins}
\usetikzlibrary{external}
\pgfplotstableread[col sep = comma]{data/HertzianContact/Hertz_analytical.csv}\HertzAnalytical
\pgfplotstableread[col sep = comma]{data/HertzianContact/hertz_rgd_sym_coll_gal/hertz_rgd_sym_coll_gal_mesh_nel_u_25_mesh_nel_v_25.csv}\HertzCollGalTwentyFive
\pgfplotstableread[col sep = comma]{data/HertzianContact/hertz_rgd_sym_coll_gal/hertz_rgd_sym_coll_gal_mesh_nel_u_50_mesh_nel_v_50.csv}\HertzCollGalFifty
\pgfplotstableread[col sep = comma]{data/HertzianContact/hertz_rgd_sym_modcoll_gal/hertz_rgd_sym_modcoll_gal_mesh_nel_u_25_mesh_nel_v_25.csv}\HertzModCollGalTwentyFive
\pgfplotstableread[col sep = comma]{data/HertzianContact/hertz_rgd_sym_modcoll_gal/hertz_rgd_sym_modcoll_gal_mesh_nel_u_50_mesh_nel_v_50.csv}\HertzModCollGalFifty
\pgfplotstableread[col sep = comma]{data/HertzianContact/hertz_rgd_sym_coll/hertz_rgd_sym_coll_mesh_nel_u_25_mesh_nel_v_25.csv}\HertzCollTwentyFive
\pgfplotstableread[col sep = comma]{data/HertzianContact/hertz_rgd_sym_coll/hertz_rgd_sym_coll_mesh_nel_u_50_mesh_nel_v_50.csv}\HertzCollFifty
\pgfplotstableread[col sep = comma]{data/HertzianContact/hertz_rgd_sym_modcoll/hertz_rgd_sym_modcoll_mesh_nel_u_25_mesh_nel_v_25.csv}\HertzModCollTwentyFive
\pgfplotstableread[col sep = comma]{data/HertzianContact/hertz_rgd_sym_modcoll/hertz_rgd_sym_modcoll_mesh_nel_u_50_mesh_nel_v_50.csv}\HertzModCollFifty
\pgfplotstableread[col sep = comma]{data/HertzianContact/hertz_rgd_sym_pure_gal/hertz_rgd_sym_pure_gal_mesh_nel_u_25_mesh_nel_v_25.csv}\HertzPureGalTwentyFive
\pgfplotstableread[col sep = comma]{data/HertzianContact/hertz_rgd_sym_pure_gal/hertz_rgd_sym_pure_gal_mesh_nel_u_50_mesh_nel_v_50.csv}\HertzPureGalFifty
\pgfplotstableread[col sep = comma]{data/HertzianContact/hertz_rgd_sym_gal_pe_pts/hertz_rgd_sym_gal_pe_pts_mesh_nel_u_25_mesh_nel_v_25.csv}\HertzGalPePtsTwentyFive
\pgfplotstableread[col sep = comma]{data/HertzianContact/hertz_rgd_sym_gal_pe_pts/hertz_rgd_sym_gal_pe_pts_mesh_nel_u_50_mesh_nel_v_50.csv}\HertzGalPePtsFifty
\pgfplotstableread[col sep = comma]{data/HertzianContact/hertz_rgd_sym_gal_pe_pts_plus/hertz_rgd_sym_gal_pe_pts_plus_mesh_nel_u_25_mesh_nel_v_25.csv}\HertzGalPePtsPlusTwentyFive
\pgfplotstableread[col sep = comma]{data/HertzianContact/hertz_rgd_sym_gal_pe_pts_plus/hertz_rgd_sym_gal_pe_pts_plus_mesh_nel_u_50_mesh_nel_v_50.csv}\HertzGalPePtsPlusFifty

\pgfplotstableread[col sep = comma]{data/Ironing/ReactionForcesHorP2.csv}\IroningHorPTwo
\pgfplotstableread[col sep = comma]{data/Ironing/ReactionForcesHorP3.csv}\IroningHorPThree
\pgfplotstableread[col sep = comma]{data/Ironing/ReactionForcesHorP4.csv}\IroningHorPFour
\pgfplotstableread[col sep = comma]{data/Ironing/ReactionForcesVertP2.csv}\IroningVertPTwo
\pgfplotstableread[col sep = comma]{data/Ironing/ReactionForcesVertP3.csv}\IroningVertPThree
\pgfplotstableread[col sep = comma]{data/Ironing/ReactionForcesVertP4.csv}\IroningVertPFour

\begin{document}

\newcolumntype{L}[1]{>{\raggedright\arraybackslash}p{#1}}
\newcolumntype{C}[1]{>{\centering\arraybackslash}p{#1}}
\newcolumntype{R}[1]{>{\raggedleft\arraybackslash}p{#1}}

\begin{frontmatter}

\title{The isogeometric collocated contact surface approach}
\author[label1]{Frederik Fahrendorf}
\ead{f.fahrendorf@tu-braunschweig.de}
\author[label2]{Laura De Lorenzis\corref{cor2}}
\ead{ldelorenzis@ethz.ch}
\address[label1]{Institute of Applied Mechanics, TU Braunschweig, \\ Pockelsstraße 3, 38106 Braunschweig, Germany}
\address[label2]{Department of Mechanical and Process Engineering, ETH Zurich, \\ Tannenstraße 3, 8092 Zurich, Switzerland}
\cortext[cor2]{Corresponding author.}

\begin{abstract}
We propose a frictionless contact formulation for isogeometric analysis, which combines a collocated formulation for the contact surfaces with a standard Galerkin treatment of the bulk. We denote it as isogeometric Collocated Contact Surface (CCS) formulation. The approach is based on a simple \emph{pointwise} enforcement of the contact constraints, performed in this study with the penalty method. Unlike pointwise (node-to-surface or point-to-surface) contact algorithms in the Galerkin framework, the CCS formulation passes the contact patch test to machine precision by naturally exploiting the favorable properties of isogeometric collocation. Compared with approaches where the discretization of both bulk and contact surfaces is based on collocation, the CCS approach does not need enhancements to remove oscillations for highly non-uniform meshes. With respect to \emph{integral} contact approaches, the CCS algorithm is less expensive, easier to code and can be added to a pre-existing isogeometric analysis code with minimal effort. Numerical examples in both small and large deformations are investigated to compare the CCS approach with some available contact formulations and to demonstrate its accuracy.
\end{abstract}


\end{frontmatter}
\section{Introduction}
\label{sec:Introduction}

The numerical simulation of contact problems is a challenging task due to the highly non-linear and non-smooth nature of contact processes. With the finite element method (FEM), the non-smooth contact surface discretization obtained with $C^0$ Lagrange basis functions leads to ill-defined normals at the inter-element boundaries, which reduce the robustness of the simulations and call for \emph{ad hoc} remedies in contact projection algorithms. To circumvent this problem, various surface smoothing algorithms have been proposed \cite{neto2017surface,de2014isogeometric}, with the main idea of replacing the surface representation by a smoother approximation, e.g. based on Hermite polynomials, Bézier curves or non-uniform rational B-Splines (NURBS), while leaving the bulk unchanged. These procedures lead in general to a more robust contact behavior, however, the design of a smoothed master surface is far from trivial, especially in the three-dimensional space, and additional complications arise in implementation and data management.

In contrast to FEM, Isogeometric Analysis (IGA) is inherently based on basis functions featuring higher order and higher smoothness, such as B-Splines or NURBS. Thus IGA naturally suppresses discretization-induced contact projection issues, which makes it particularly appealing for computational contact mechanics. As a result, isogeometric contact attracted significant attention, see \cite{de2014isogeometric} for a review. The same paper introduces the terminology that we adopt in the present paper for the different considered types of contact formulations. 

Contact discretization schemes in either FEM or IGA can be roughly classified into \emph{pointwise} and \emph{integral} approaches. In the \emph{pointwise} category, the integral representing the contact contribution to the weak form is computed through point collocation at a given set of points, typically slave nodes (in FEM) or the physical maps of the knots on the slave surface (in IGA). The classical example of this category is the “node-to-surface” or (in two dimensions) “node-to-segment” (NTS) algorithm, which in the isogeometric context is more appropriately denoted as “point-to-surface” or “point-to-segment” (PTS) \cite{matzen2013point}. A modified version with weighted contributions was developed in \cite{matzen2016weighted} and denoted as PTS+, and this is the pointwise approach that we consider in the present paper for comparison. The main advantages of pointwise methods are their simplicity and computational efficiency, whereas their most evident drawback is their inability to pass the contact patch test \cite{taylor1991patch}, which implies that the discretization error may not decrease with decreasing mesh size. On the other hand, in \emph{integral} approaches the contact contribution to the weak form is computed through different types of integration. These approaches pass the patch test, either up to the integration error or to machine precision, however their better accuracy goes at the expenses of implementational simplicity and computational efficiency. In this paper, we adopt for comparison what we consider to be the simplest available version of integral approaches, namely the “Gauss-point-to-segment” (GPTS) formulation, first proposed in the isogeometric context by \cite{temizer2011contact, de2011large, dimitri2014nurbs} in conjunction with the penalty method for constraint enforcement. We do not compare here with the mortar method, which is the best performing, but also most complex and computationally expensive integral approach \cite{temizer2011contact,de2011large,temizer2012three}.

Typically, both pointwise and integral contact discretizations introduce a bias between slave and master surfaces. Due to the use of a single loop over the slave mesh features at which the contact constraints are enforced, these formulations are also referred to as “one-pass”. With the purpose of eliminating the bias, “two-pass” formulations have been proposed, where the contact contribution to the weak form is incorporated twice while switching the roles of slave and master surfaces. Obviously, the number of constraints increases, which in general leads to overconstraining (or worsens possible pre-existing overconstraining of the one-pass counterpart). For the NTS discretization, the two-pass version in combination with the Lagrange multiplier method passes the contact patch test \cite{taylor1991patch}, whereas the same version with the penalty method still fails it. An alternative to one-pass and two-pass formulations was explored in \cite{papadopoulos1995novel} and later recovered (with some differences) in \cite{sauer2013computational} from a very general framework based on surface potentials. In this approach, two loops are performed treating each surface alternatively as slave and master. In each loop (“half-pass”), the contact tractions are computed only on the surface currently treated as slave. Therefore, no transfer of tractions to the master side is needed. Local equilibrium at the surfaces is not enforced \emph{a priori} but recovered with high accuracy. The approach, denoted in \cite{sauer2013computational,sauer2015unbiased} as “two-half-pass”, was applied in combination with the NTS \cite{papadopoulos1995novel} and GPTS discretizations \cite{sauer2013computational,sauer2015unbiased}. As shown in \cite{sauer2015unbiased}, the GPTS formulation in conjunction with the two-half-pass algorithm and the penalty method passes the contact patch test to machine precision. Applications of two-half-pass approaches within an isogeometric framework can be found e.g. in \cite{lu2011isogeometric,duong2017efficient,duong2019concise,duong2019segmentation}. Here overconstraining is still an issue, unless special precautions such as node patterning are taken \cite{papadopoulos_solberg}. Obviously, the issue is less severe with the penalty method. In this paper, the two-half-pass approach is combined with both the PTS+ and GPTS formulations. 

The contact formulations that we briefly summarized above are all based on the computation of the contact contribution to the weak form, i.e. they are rooted in Galerkin-based FEM or IGA. Unlike IGA, isogeometric collocation starts directly from the strong form of the governing equations, which is enforced at a set of evaluation points equal in number to control points and denoted as collocation points \cite{reali2015introduction,auricchio2010isogeometric,auricchio2012isogeometric}. Thus there is no need for numerical quadrature of integral equations and the cost of assembly is minimized, leading to a high computational efficiency especially for higher-order discretizations  \cite{schillinger2013isogeometric}. Contact formulations for isogeometric collocation were developed for linear elasticity without friction \cite{de2015isogeometric}, hyperelasticity with friction \cite{kruse2015isogeometric} and Cosserat rods with friction \cite{weeger2017isogeometric,weeger2018isogeometric} in combination with the penalty method. The contact treatment in \cite{de2015isogeometric,kruse2015isogeometric} is based on the strong enforcement of the contact constraints at the collocation points located on the contact surfaces, hence it is a pointwise approach; nevertheless, it passes the contact patch test to machine precision. The reason is that in the framework of isogeometric collocation the governing equations naturally involve contact pressures, as opposed to the concentrated nodal forces of Galerkin-based formulations. Moreover, the contact formulation in \cite{de2015isogeometric,kruse2015isogeometric} is based on the two-half-pass approach, which naturally fits the framework of isogeometric collocation. 

The contact formulation in \cite{de2015isogeometric,kruse2015isogeometric} was developed for a framework where both bulk and contact are treated with isogeometric collocation. As a result, while enjoying computational efficiency and inherent contact patch test satisfaction, it also suffered from the known drawbacks of isogeometric collocation, i.e. stress oscillations at the Neumann (and at the contact) boundary for highly non-uniform meshes, which were solved using the enhanced collocation approach \cite{de2015isogeometric}. Moreover, isogeometric collocation as a bulk discretization method requires sufficient regularity of the material behavior if a primal formulation is to be used, e.g. its application to $J_2$ plasticity requires mixed formulations \cite{fahrendorf2020mixed}, and may need special attention to achieve good robustness in certain cases \cite{fahrendorf2020mixed,fahrendorf2021mixed}. 

In this work we propose a novel approach, in which the standard IGA formulation for the bulk is combined with a contact formulation based on isogeometric collocation. We denote it as isogeometric Collocated Contact Surface (CCS) approach. It is based on a simple \emph{pointwise} enforcement of the contact constraints, performed in this study with the penalty method. Unlike pointwise contact algorithms in the Galerkin framework, the CCS formulation passes the contact patch test to machine precision by naturally exploiting the favorable properties of isogeometric collocation. Compared with approaches where the discretization of both bulk and contact surfaces is based on collocation, the CCS approach does not need enhancements to remove oscillations for highly non-uniform meshes. Moreover, it enjoys the flexibility  and robustness of the Galerkin framework in the bulk discretization. With respect to \emph{integral} contact approaches, the CCS algorithm is less expensive and easier to code, and can be added to a pre-existing isogeometric analysis code with minimal effort. 
In this paper, we focus on the two-dimensional frictionless setting. 

The paper is structured as follows. In Section \ref{sec:ElastostaticContactProblem}, we formulate the elastostatic boundary value problem with frictionless contact in strong and weak form in the continuum setting. Section \ref{sec:IsogeometricGalerkinCollocationMethods} reviews isogeometric Galerkin and collocation methods for bulk and contact discretization, including some available contact formulations. Section \ref{sec:CollocatedContactSurfaceApproach} introduces the newly proposed isogeometric CCS approach. The performance of CCS and available formulations is compared by means of numerical examples in Section \ref{sec:NumericalExamples}. Conclusions are drawn in Section \ref{sec:SummaryConclusions}.

\section{Elastostatic problem with frictionless contact}
\label{sec:ElastostaticContactProblem}

Contact processes are typically associated with large deformations of the considered continua. As follows, we outline the fundamental continuum equations of elastostatics with contact in the finite strain setting.    

\subsection{Strain and stress measures, constitutive laws}
\label{subsec:Kinematics}

Let us start by considering a single continuum body $\mathcal{B}$ undergoing finite deformations. The undeformed configuration is parameterized with the reference coordinates $\boldsymbol{X}$ and the deformed configuration with the current coordinates $\boldsymbol{x}$, with the deformation defined by the mapping $\boldsymbol{x}=\pmb\varphi\left(\boldsymbol{X},t\right)$. Accordingly the displacements are defined as $\boldsymbol{u}=\boldsymbol{x}-\boldsymbol{X}$. The domains occupied by $\mathcal{B}$ in the reference and the current configuration are denoted as $\Omega, \omega \subset\mathbb{R}^{d}$, respectively, with $d$ as the number of space dimensions ($d=2$ in this paper).

The deformation gradient is defined as $\boldsymbol{F}=\boldsymbol{I} + \nabla \boldsymbol{u}$, where $\nabla$ is the gradient operator with respect to $\boldsymbol{X}$ and $\boldsymbol{I}$ is the second-order identity tensor. The Jacobian $J = \text{det} \boldsymbol{F}$ is a measure for the transformation of volume elements between the two considered configurations. We adopt as strain measure the Green-Lagrange strain tensor $\boldsymbol{E} = \frac{1}{2}(\boldsymbol{C}-\boldsymbol{I})$, with $\boldsymbol{C} = \boldsymbol{F}^T \boldsymbol{F}$ known as the right Cauchy-Green deformation tensor.

For numerical examples exhibiting large deformations, we consider a hyperelastic neo-Hookean material model \cite{simo1984remarks} for which the elastic strain energy density $\psi(\boldsymbol{C})$  reads
\begin{equation}
\psi=\frac{\mu}{2}(I_{1}-3)-\mu\text{ln}J+\frac{\lambda}{2}(\text{ln}J)^{2},
\end{equation}
where $\lambda$ and $\mu$ are the Lam\'e parameters and $I_{1}=\text{tr}\mathbf{C}=\mathbf{C}:\mathbf{I}$ is the first invariant of $\mathbf{C}$. 

The second Piola–Kirchhoff stress tensor $\boldsymbol{S}$ can be obtained as 
\begin{equation}
\mathbf{S}=\frac{\partial \psi(\boldsymbol{C})}{\partial \boldsymbol{E}} =2 \frac{\partial \psi(\boldsymbol{C})}{\partial \boldsymbol{C}} = \mu(\mathbf{I}-\mathbf{C}^{-1})+\lambda\text{ln}J\mathbf{C}^{-1}.
\end{equation}
The first Piola-Kirchhoff stress tensor $\boldsymbol{P}$ follows through the relation $\boldsymbol{P} = \boldsymbol{F} \boldsymbol{S}$. The Piola traction vector $\boldsymbol{T}$ can be calculated as $\boldsymbol{T}= \boldsymbol{P} \boldsymbol{N}$ with the outward unit normal $\boldsymbol{N}$ to a surface element in the reference configuration. The relation to the traction vector $\boldsymbol{t}$ in the current configuration, which is an important quantity in contact formulations, is given by
\begin{equation}
\boldsymbol{t}(\boldsymbol{x},\boldsymbol{n})da = \boldsymbol{T}(\boldsymbol{X},\boldsymbol{N})dA
\label{Nanson}
\end{equation}
with the infinitesimal surface elements $da$ and $dA$ in the current and in the reference configuration, respectively, and with $\boldsymbol{n}$ as the outward unit normal to a surface element in the current configuration. 

In the special case of small deformations, a distinction between the reference and the current configuration is not necessary and the linearized strain tensor $\boldsymbol{\varepsilon}^{lin}=\frac{1}{2}(\nabla \boldsymbol{u} + (\nabla \boldsymbol{u})^T)$ can be used as strain measure. For linearly elastic isotropic materials, the Cauchy stress tensor $\boldsymbol{\sigma}^{lin}$ and the linearized strain tensor $\boldsymbol{\varepsilon}^{lin}$ are related by Hooke’s law as $\boldsymbol{\sigma}^{lin}=(\lambda\mathbf{I}\otimes\mathbf{I}+2\mu\mathbf{\Pi}):\boldsymbol{\varepsilon}^{lin}$, where $\mathbf{\Pi}$ is the fourth-order symmetric identity tensor.

\subsection{Contact formulation in the continuum setting}
\label{subsec:ContactKinematics}

Let us now consider two elastic bodies $\mathcal{B}^{(k)}$ ($k = {1,2}$) that come into contact under the assumption of large deformations. For both bodies $\boldsymbol{x}^{(k)} = \boldsymbol{X}^{(k)} + \boldsymbol{u}^{(k)}$ holds.

The contact surface of body $\mathcal{B}^{(k)}$ in the current configuration, $\gamma_C^{(k)}$, is parameterized via the convective coordinates $\xi^{\alpha (k)}$, $\alpha \in {1,..,d-1}$, that define the covariant (tangent) vectors $\boldsymbol{\tau}^{k}_\alpha = \boldsymbol{x}^{(k)}_{,\alpha}$. Based on the tangent vectors we can further define $\boldsymbol{n}^{(k)}$ as the outward normal unit vector.

The closest-point (normal) projection of a given point $\boldsymbol{x}^{(1)}$ of surface $\gamma_C^{(1)}$ onto the matching surface $\gamma_C^{(2)}=\gamma_C^{(1)}=\gamma_C$ identifies the projection point $\bar{\boldsymbol{x}}^{(2)}$. Thus the normal gap $g^{(2)}_N$ can be computed as
\begin{equation}
g^{(2)}_N = (\boldsymbol{x}^{(1)} - \boldsymbol{\bar{x}}^{(2)})\cdot \bar{\boldsymbol{n}}^{(2)} 
\label{eq:GapFunction}
\end{equation}
where $\bar{\boldsymbol{n}}^{(2)}$ denotes the normal to $\gamma_C^{(2)}$ at the projection point. Indicating as $\boldsymbol{t}^{(2)}$ the contact traction vector acting on surface $\gamma_C^{(2)}$, and with $\boldsymbol{t}^{(2)}_N$ its component in the direction of $\bar{\boldsymbol{n}}^{(2)}$, it is for frictionless contact
\begin{equation}
\boldsymbol{t}^{(2)}=\boldsymbol{t}^{(2)}_N = t^{(2)}_N \bar{\boldsymbol{n}}^{(2)}
\label{eq:NormalTraction}
\end{equation}
with $t^{(2)}_N$ as the normal component of the traction vector. 
If the two bodies are in contact at the considered point, it is $g^{(2)}_N = 0$ and $t^{(2)}_N \leq 0$. If the contact is open, it is $g^{(2)}_N \geq 0$ and $t^{(2)}_N = 0$. Thus the contact constraints can be formulated as the following Karush-Kuhn-Tucker (or Hertz-Signorini-Moreau) conditions
\begin{equation}
\label{KKT}
g^{(2)}_N \geq 0, \quad t^{(2)}_N \leq 0, \quad g^{(2)}_N t^{(2)}_N = 0.
\end{equation}
The computation of the contact traction depends on the solution method chosen for the enforcement of the contact constraints. Here we adopt the penalty method. The penalty regularized contact constraints read
\begin{equation}
t^{(2)}_N = \epsilon_N \langle g^{(2)}_N \rangle_{-} \quad \langle g^{(2)}_N \rangle_{-}\begin{cases} g^{(2)}_N & \text{if } g^{(2)}_N \leq 0, \\ 0 & \text{otherwise,} \end{cases}
\label{penalty} 
\end{equation} 
where $\epsilon_N > 0$ is the so-called penalty parameter and $\langle \bullet \rangle_{-}$ denotes the Macaulay brackets. Thus the penalty approach regularizes the contact constraints in (\ref{KKT}) by allowing for a small penetration of the contacting bodies.

\begin{figure}[htb]
\centering
\includegraphics[width=0.4\textwidth]{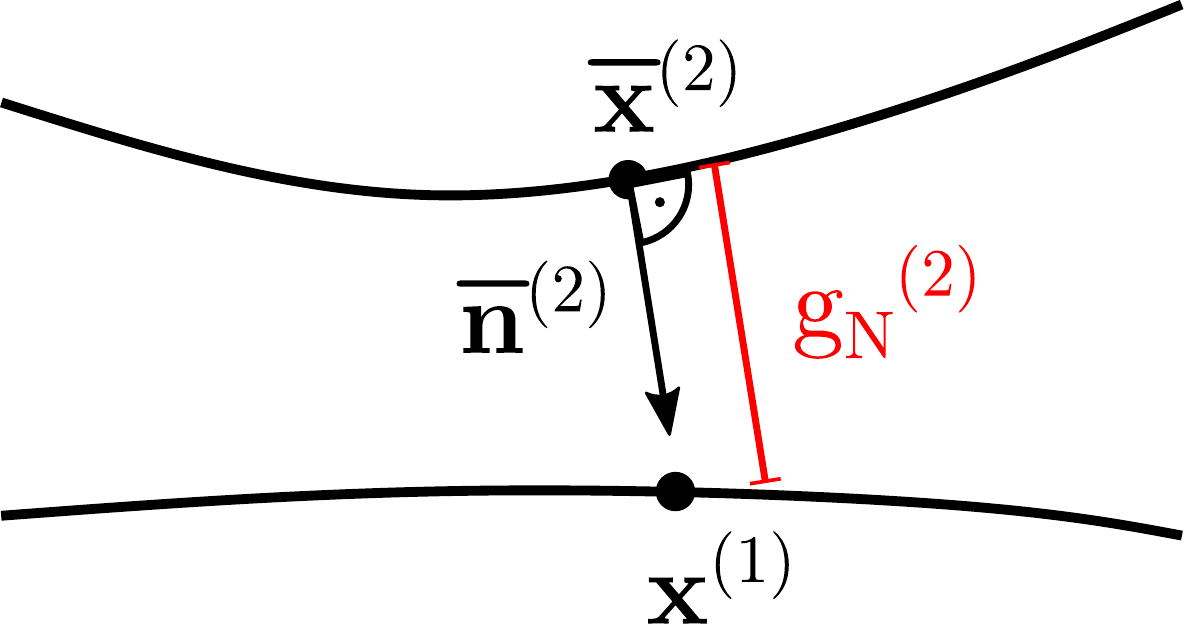}
\caption{Schematic representation of contact kinematics between two bodies.}
\label{fig:ContactKinematics}
\end{figure}

In one-pass approaches, body $\mathcal{B}^{(1)}$ is treated as slave and body $\mathcal{B}^{(2)}$ as master, the contact traction on the master body is computed from (\ref{eq:NormalTraction}) and (\ref{penalty}), whereas the contact traction on the slave body follows from the action-reaction principle as
\begin{equation}
\boldsymbol{t}^{(1)} = -\boldsymbol{t}^{(2)}.
\label{eq:PiolaTractionContact}
\end{equation}
Instead, in two-half-pass formulations, two loops are performed treating each surface alternatively as slave and master. In each loop, the contact tractions are computed only on the surface currently treated as slave. Thus, in addition to the ones introduced previously, the following symmetric relationships are used
\begin{equation}
g_N^{(1)} = (\boldsymbol{x}^{(2)} - \boldsymbol{\bar{x}}^{(1)})\cdot \bar{\boldsymbol{n}}^{(1)}
\label{eq:TwoHalfPassGapFunction}
\end{equation}
\begin{equation}
\boldsymbol{t}^{(1)}=\boldsymbol{t}_N^{(1)} = t_N^{(1)} \bar{\boldsymbol{n}}^{(1)}
\label{eq:TwoHalfPassNormalTraction}
\end{equation}
%
%
\begin{equation}
t_N^{(1)}= \epsilon_N \langle g_N^{(1)} \rangle_{-} \qquad \langle g_N^{(1)} \rangle_{-}\begin{cases} g_N^{(1)} & \text{if } g_N^{(1)}\leq 0, \\ 0 & \text{otherwise,} \end{cases}
\label{eq:TwoHalfPassPenaltyNormalTraction}
\end{equation}
where the contact constraints are given directly after penalty regularization. Notice that now (\ref{eq:PiolaTractionContact}) is no longer needed. Thus, equilibrium at the contact surfaces is no longer explicitly enforced but recovered \emph{a posteriori} with high accuracy \cite{sauer2013computational,sauer2015unbiased}.
\subsection{Boundary value problem in strong form}
\label{subsec:BVPStrongForm}
The elastostatic boundary value problem with contact for the two bodies $\mathcal{B}^{(k)}$ occupying the domains $\Omega^{(k)}$ in the reference configuration is formulated in the following in strong form. Each boundary $\Gamma^{(k)}$  can be subdivided into a portion $\Gamma_D^{(k)}$ with Dirichlet boundary conditions, a portion $\Gamma_N^{(k)}$ with Neumann boundary conditions, and the remaining portion $\Gamma_C^{(k)}$ on which contact constraints hold, with $\Gamma^{(k)} = \Gamma_D^{(k)}\bigcup \Gamma_N^{(k)}\bigcup \Gamma_C^{(k)}$ and $\Gamma_D^{(k)}\bigcap \Gamma_N^{(k)} = \Gamma_N^{(k)} \bigcap \Gamma_C^{(k)} = \Gamma_C^{(k)} \bigcap \Gamma_D^{(k)} = \emptyset$. 

The balance of linear momentum reads 
\begin{equation}
\nabla \cdot \boldsymbol{P}^{(k)} + \boldsymbol{B}^{(k)} = \boldsymbol{0} \qquad \text{in }\Omega^{(k)}
\end{equation}
with the first Piola-Kirchhoff stress tensors $\boldsymbol{P}^{(k)}$ and the body forces $\boldsymbol{B}^{(k)}$. 
On the Dirichlet boundaries $\Gamma^{(k)}_D$ displacements $\boldsymbol{\bar{u}}^{(k)}$ are prescribed
\begin{equation}
\boldsymbol{u}^{(k)} = \boldsymbol{\bar{u}}^{(k)} \qquad \text{on }\Gamma_D^{(k)}
\end{equation}
whereas tractions $\boldsymbol{\bar{T}}$ are applied to the Neumann boundaries $\Gamma^{(k)}_{N}$
\begin{equation}
\boldsymbol{P}^{(k)} \boldsymbol{N}^{(k)} = \boldsymbol{\bar{T}}^{(k)} \qquad \text{on }\Gamma_N^{(k)}
\end{equation}
with $\boldsymbol{N}^{(k)}$ as the outward normal unit vector to $\Gamma^{(k)}_{N}$ in the reference configuration.
On the contact surfaces it is
\begin{equation}
\boldsymbol{P}^{(k)} \boldsymbol{N}^{(k)} = \boldsymbol{T}^{(k)} \qquad \text{on }\Gamma_C^{(k)}.
\end{equation}
Here $\boldsymbol{T}^{(k)}$ denotes the contact Piola traction vectors, which are computed from the contact Cauchy traction vectors $\boldsymbol{t}^{(k)}$ of Section \ref{subsec:ContactKinematics} by accounting through (\ref{Nanson}) for the mapping from the current to the reference configuration. 

\subsection{Variational formulation}
\label{subsec:BVPVariationalFormulation}

Based on the principle of virtual work, the finite deformation elasticity problem in variational form, expressed in the reference configuration, consists of finding $\boldsymbol{u}^{(k)} \in \boldsymbol{\mathcal{U}}^{(k)}$ such that for all $\delta \boldsymbol{u}^{(k)} \in \boldsymbol{\mathcal{V}}^{(k)}$
\begin{equation}
\begin{split}
\sum_{k=1}^2 & \left[\underbrace{\int_{\Omega^{(k)}}\boldsymbol{P}^{(k)}:\nabla \delta \boldsymbol{u}^{(k)}\,d\Omega}_{\delta W_{int}^{(k)}} \underbrace{-\int_{\Omega^{(k)}}\boldsymbol{B}^{(k)}\cdot\delta \boldsymbol{u}^{(k)}\,d\Omega - \int_{\Gamma_{N}^{(k)}}\boldsymbol{\bar{T}}^{(k)}\cdot\delta \boldsymbol{u}^{(k)}\,d\Gamma}_{\delta W_{ext}^{(k)}} \right. \\
  &\left. \underbrace{-\int_{\Gamma_{C}^{(k)}}\boldsymbol{T}^{(k)}\cdot\delta \boldsymbol{u}^{(k)}\,d\Gamma}_{\delta W_{C}^{(k)}}\right] = 0
\end{split}
\label{eq:BVPVariationalFormulation}
\end{equation}
with the following definition for the approximation spaces
\begin{equation}
\begin{split}
&\boldsymbol{\mathcal{U}}^{(k)} = \{\boldsymbol{u}^{(k)}|\boldsymbol{u}^{(k)}\text{ suff. regular},\, \boldsymbol{u}^{(k)}|_{\Gamma^{(k)}_D} = \boldsymbol{\bar{u}}^{(k)}\}, \\
&\boldsymbol{\mathcal{V}}^{(k)} = \{\delta \boldsymbol{u}|\delta \boldsymbol{u}^{(k)}\text{ suff. regular},\, \delta \boldsymbol{u}^{(k)}|_{\Gamma^{(k)}_D} = \boldsymbol{0}\}.
\end{split}
\end{equation}
where we do not further specify the regularity requirements here. Let us now focus on the contact contribution to the weak form,
\begin{equation}
\delta W_{C}=\sum_{k=1}^{2}\delta W_{C}^{(k)}=-\sum_{k=1}^{2}\int_{\Gamma_{C}^{(k)}}\mathbf{T}^{(k)}\cdot\delta\mathbf{u}^{(k)}d\Gamma
\end{equation}
In alternative than in the reference configuration, where in general $\Gamma_{C}^{(1)}\neq\Gamma_{C}^{(2)}$, $\delta W_{C}$
can also be expressed in the current configuration as follows
\begin{equation}
\delta W_C = -\int_{\gamma_C} \boldsymbol{t}^{(2)} \cdot  \delta \bar{\boldsymbol{x}}^{(2)}  d\gamma - \int_{\gamma_C} \boldsymbol{t}^{(1)}  \cdot \delta \boldsymbol{x}^{(1)} d\gamma
\label{eq:concurr}
\end{equation}
which exploits the coincidence of the contact surfaces in the current configuration noted earlier, $\gamma_C^{(1)}=\gamma_C^{(2)}=\gamma_C$, and their pairing through closest-point projection.

Considering a conventional master-slave treatment of the contact surface, this expression can be further simplified using Eqs.~\eqref{eq:PiolaTractionContact},\eqref{eq:NormalTraction},\eqref{eq:GapFunction} and the variation of the gap function $\delta g^{(2)}_N = (\delta \boldsymbol{x}^1 - \delta \bar{\boldsymbol{x}}^2)\cdot \bar{\boldsymbol{n}}^{(2)}$, resulting in 
\begin{equation}
\delta W_C = \int_{\gamma_{C}} t_N \delta g_N d\gamma  = \epsilon_N \int_{\gamma_{C}} g_N \delta g_N  d\gamma.
\label{eq:VirtualWork_MS}
\end{equation}
where we have used \eqref{penalty}, defined $t_N=t^{(2)}_N$ and $g_N=g^{(2)}_N$, and removed the Macauley brackets under the assumption to have identified the (active) contact surface using a suitable active set strategy. Instead, with a two-half-pass treatment, the contact virtual work reads
\begin{equation}
\delta W_C = - \epsilon_N \int_{\gamma_{C}} g_N^{(1)} \boldsymbol{n}^{(1)} \cdot  \delta \boldsymbol{x}^{(1)}  d\gamma - \epsilon_N \int_{\gamma_{C}} g_N^{(2)}  \boldsymbol{n}^{(2)} \cdot \delta \boldsymbol{x}^{(2)}   d\gamma.
\label{eq:VirtualWork_THP}
\end{equation}
which coincides with \eqref{eq:concurr} combined with penalty regularization.

\section{Isogeometric Galerkin and collocation methods}
\label{sec:IsogeometricGalerkinCollocationMethods}

In this section, we first review the basics of B-spline and NURBS basis functions. Then we briefly illustrate isogeometric Galerkin and collocation methods including both bulk and contact discretization.


\subsection{B-Spline and NURBS basis functions, collocation points}
\label{subsec:Nurbs}

A B-spline basis of degree $p$ is constructed based on a so-called knot vector, i.e. a non-decreasing sequence of real numbers 
$\boldsymbol{\Xi}=\{\xi_1,\xi_2,\dots,\xi_{n+p+1}\}$, where each $\xi_i$ is a knot and $n$ denotes the number of basis functions of degree $p$. Throughout this paper, the knot vector is assumed to be open, which implies $\xi_1 = \ldots = \xi_{p+1}$ and $\xi_{n+1} = \ldots =\xi_{n+p+1}$. If a knot has multiplicity $k$, the continuity of the B-spline basis is $C^{p-k}$ at that knot. The continuity is $C^\infty$ in the interior of a knot span.

A common choice for the location of collocation points in isogeometric collocation are the Greville abscissae of the knot vectors. For a B-Spline basis of degree $p$ the Greville abscissae are defined  as
\begin{equation}
\hat{\tau}_i = \frac{1}{p} \sum_{j=i+1}^{i+p} \xi_j, \;\;\; i=1,...,n
\label{eq:GrevilleAbscissae}
\end{equation}
For multivariate discretizations, the Greville abscissae are obtained via the tensor product of \eqref{eq:GrevilleAbscissae} in the various parametric directions. 

The univariate $p$-th degree B-Spline basis functions $\{N_{i,p}\}_{i=1,\dots,n}$ are defined by means of the Cox-de Boor recursion formula using the relations
\begin{subequations}
\begin{alignat}{2}
N_{i,0}(\xi)&=\begin{cases}
	1\quad \text{if}\quad \xi_i\leq\xi <  \xi_{i+1}, \\
	0\quad \text{otherwise,}
\end{cases}\\
	N_{i,p}(\xi) &= \frac{\xi-\xi_i}{\xi_{i+p}-\xi_i}N_{i,p-1}(\xi) + \frac{\xi_{i+p+1}-\xi}{\xi_{i+p+1}-\xi_{i+1}}N_{i+1,p-1}(\xi).
\end{alignat}
\end{subequations}
and adopting the convention $\frac{0}{0} = 0$. 

Bivariate NURBS basis functions $R_{i,j}$ of degrees $p$ and $q$ in the two parametric directions $\xi$ and $\eta$ with the corresponding weights $w_{i,j}$ are defined by a product of the univariate B-spline basis functions $N_{i,p}(\xi)$, $M_{j,q}(\eta)$ as
\begin{equation}
R_{i,j}(\xi,\eta) = \frac{N_{i,p}(\xi)M_{j,q}(\eta)w_{i,j}}{\sum_{\hat{i}=1}^n \sum_{\hat{j}=1}^m N_{\hat{i},p}(\xi)M_{\hat{j},q}(\eta)w_{\hat{i},\hat{j}}}.
\end{equation} 
A NURBS surface of degree $p,q$ can be expressed as a linear combination of control points $\boldsymbol{P}_{i,j}$ with the corresponding basis functions $R_{i,j}$ as 
\begin{equation}
\boldsymbol{S}(\xi,\eta) = \sum_{i=1}^n \sum_{j=1}^{m} R_{i,j}(\xi,\eta) \boldsymbol{P}_{i,j}.
\label{eq:nurbssurf}
\end{equation}

\subsection{Galerkin formulation}
\label{subsec:GalerkinBulk}

IGA, like FEM, is based on the discretization of the weak form (\ref{eq:BVPVariationalFormulation}). The unknown displacement fields $\boldsymbol{u}^{(k)}$ are approximated as follows
\begin{equation}
\boldsymbol{u}^{(k)}\approx\boldsymbol{u}^{(k) h}=\sum_{a=1}^{N^{(k)}}R_{a}\hat{\boldsymbol{u}}^{(k)}_{a},\label{eq:DiscretizationBasisFunctions}
\end{equation}
where $R_{a}$ are NURBS basis functions in IGA (as opposed to the Lagrange basis functions used in FEM) and $\hat{\boldsymbol{u}}^{(k)}_{a}$ are the $N^{(k)} = n^{(k)}m^{(k)}$ unknown displacement control variables of body $\mathcal{B}^{(k)}$. The symbol $(\bullet)^h$ indicates discretized quantities. Note that, for convenience, we have summarized the two indices $i,j$ of \eqref{eq:nurbssurf} in a single running index $a$, with $a = (j-1)n + i$. According to the Bubnov-Galerkin approach, the test functions (or virtual displacements) $\delta \boldsymbol{u}^{(k)}$ are discretized with the same ansatz: 
\begin{equation}
\delta \boldsymbol{u}^{(k)} \approx \delta \boldsymbol{u}^{(k) h}=\sum_{a=1}^{N^{(k)}}R_{a}\delta\hat{\boldsymbol{u}}^{(k)}_{a}.\label{eq:DiscretizationTestFunctions}
\end{equation}
Upon substitution of (\ref{eq:DiscretizationBasisFunctions}) and (\ref{eq:DiscretizationTestFunctions}) in \eqref{eq:BVPVariationalFormulation}, we obtain the IGA Galerkin formulation 
\begin{equation}
\sum_{k=1}^2 \left[\int_{\Omega^{(k)h}}\boldsymbol{P}^{(k) h}:\nabla \delta \boldsymbol{u}^{(k) h}\,d\Omega -\int_{\Omega^{(k)h}}\boldsymbol{B}^{(k)}\cdot\delta \boldsymbol{u}^{(k) h}\,d\Omega - \int_{\Gamma_{N}^{(k)h}}\boldsymbol{\bar{T}}^{(k)}\cdot\delta \boldsymbol{u}^{(k) h}\,d\Gamma \right] + \delta W_{C}^{h} = 0
\label{eq:discr_weak_form}
\end{equation}
with $\delta W_{C}^{h} $ as the contact contribution to the discretized weak form. 
The Dirichlet boundary conditions on $\Gamma^{(k)}_D$ are enforced strongly in the final system of algebraic equations.  


Let us now focus on $\delta W_{C}^{h} $. The main difference between GPTS and PTS strategies  is the choice of the quadrature rule. In GPTS, the integral(s) in Eq. \eqref{eq:VirtualWork_MS} (for the standard master-slave treatment) or \eqref{eq:VirtualWork_THP} (within a two-half-pass treatment) is/are computed with a standard Gaussian quadrature rule. This makes the approach easy to implement, but leads to a higher amount of evaluation points compared to collocation (pointwise) strategies. The contact patch test is satisfied up to the integration error for the standard master-slave, and to machine precision for the two-half-pass treatment.

PTS can be seen as GPTS with a reduced quadrature strategy. Instead of standard Gauss points, the quadrature points are here the Greville, Demko or Botella abscissae, which are equal in number to the control points. Hence, the amount of contact evaluations is reduced significantly, especially for higher-order discretizations. In the following we consider the Greville abscissae, since they coincide with the locations of the collocation points in the collocation-based contact approaches  (see Section \ref{sec:CollocatedContactSurfaceApproach}). 

In the original paper on PTS  \cite{matzen2013point}, the quadrature weights were taken equal to the unity. An improved version based on weighted contributions and denoted as PTS+ was introduced in \cite{matzen2016weighted} and is adopted here (although we still refer to it as PTS for simplicity). Quadrature weights are computed by solving the following moment-fitting system of equations   
\begin{equation}
\underbrace{
\begin{bmatrix}
           \int_{\hat{\Omega}}N_1(\xi)d\xi \\
           \int_{\hat{\Omega}}N_2(\xi)d\xi \\
           \vdots \\
           \int_{\hat{\Omega}}N_{n}(\xi)d\xi
         \end{bmatrix}
}_{\boldsymbol{F^s_{C}}}
=
\underbrace{
\begin{bmatrix}
    N_1({\hat\tau}_1) & N_1({\hat\tau}_2) & N_1({\hat\tau}_3) & \dots  & N_1({\hat\tau}_{n}) \\
    N_2({\hat\tau}_1) & N_2({\hat\tau}_2) & N_2({\hat\tau}_3)  & \dots  & N_2({\hat\tau}_{n}) \\
    \vdots & \vdots & \vdots & \ddots & \vdots \\
 N_{n}({\hat\tau}_1) & N_{n}({\hat\tau}_2) & N_{n}({\hat\tau}_3)  & \dots  & N_{n}({\hat\tau}_{n})
\end{bmatrix}
}_{\boldsymbol{G^s_{mat}}}
\underbrace{
\begin{bmatrix}
    \hat{\omega}_1 \\
    \hat{\omega}_2   \\
    \vdots   \\
    \hat{\omega}_{n}   
\end{bmatrix}    
}_{\hat{\boldsymbol{\omega}}}
\end{equation}
with the univariate B-Spline basis functions $N_i$ (of the slave contact surface) and the corresponding collocation points ${\hat\tau}_i$. The left-hand side contains the moments $\boldsymbol{F^s_{C}}$, which are computed exactly using full Gauss quadrature, and the right-hand side the unknown weights $\hat{\boldsymbol{\omega}}$ and the basis function evaluations (at the corresponding collocation points) stored in $\boldsymbol{G^s_{mat}}$. The quadrature weights are computed once at the beginning of the simulation by solving the linear system of equations for the unknown weights $\hat{\boldsymbol{\omega}}$.  

In the original PTS / PTS+ approaches, the non-penetration condition in normal direction was enforced by the Lagrange multiplier method. Here, we combine the PTS+ approach with the penalty method to allow for a better comparison with the other contact formulations.

\subsection{Isogeometric collocation}
\label{subsec:IsogeometicCollocation}

Unlike Galerkin formulations, isogeometric collocation approaches are based on solving the strong form of the boundary value problem, which is enforced at the chosen collocation points. As shown in \cite{auricchio2010isogeometric,auricchio2012isogeometric}, isogeometric collocation can alternatively be introduced based on the variational formulation (\ref{eq:BVPVariationalFormulation}) upon integration by parts and discretization (assuming sufficient regularity, which can be achieved with isogeometric basis functions) 

\begin{equation}
\begin{split}
\sum_{k=1}^2 & \left[
\int_{\Omega^{(k)}}\left[\nabla \cdot\boldsymbol{P}^{(k)}+\boldsymbol{B}^{(k)}\right]\cdot \delta\boldsymbol{u}^{(k)}\,d\Omega-\int_{{{\Gamma_{N}^{(k)}}}}\left[\boldsymbol{P}^{(k)}\boldsymbol{N}^{(k)}-\boldsymbol{\bar{T}}^{(k)}\right]\cdot \delta\boldsymbol{u}^{(k)}\,d\Gamma \right. \\
& \left. -\int_{\Gamma_{C}^{(k)}} \left[\boldsymbol{P}^{(k)}\boldsymbol{N}^{(k)}-\boldsymbol{T}^{(k)}\right]\cdot\delta \boldsymbol{u}^{(k)}\,d\Gamma \right] =0.
\end{split}
\label{eq:weighted_residual}
\end{equation}
which leads to the so-called weighted residual formulation (note that here and in the following we omit the superscript $h$ for notational simplicity). The next step is to choose for the test functions $\delta\boldsymbol{u}^{(k)}$ no longer \eqref{eq:DiscretizationTestFunctions} but the Dirac delta distribution $\delta^D$, which can be formally constructed as the limit of a sequence of smooth functions with compact support that converge to a distribution \cite{auricchio2010isogeometric,auricchio2012isogeometric}, and which satisfies the so-called sifting property, i.e.,
\begin{equation}
\int_\Omega f_\Omega(\boldsymbol{X})\delta^D(\boldsymbol{X}-\boldsymbol{X}_{i})\,d\Omega = f_\Omega(\boldsymbol{X}_{i}), \quad \int_\Gamma f_\Gamma(\boldsymbol{X})\delta^D(\boldsymbol{X}-\boldsymbol{X}_{i})\,d\Gamma = f_\Gamma(\boldsymbol{X}_{i})
\label{eq:sifting_property} 
\end{equation}
for every function $f_\Omega$ continuous about the point $\boldsymbol{X}_{i} \in \Omega$ and for every function $f_\Gamma$ continuous about the point $\boldsymbol{X}_{i} \in \Gamma$ \cite{reali2015introduction,auricchio2010isogeometric,auricchio2012isogeometric}. In the following, the Dirac delta is indicated as Dirac delta “function” following conventional terminology. The collocation points in parametric coordinates are denoted as $\boldsymbol{\hat{\tau}}_{ij}$, $i=\left\{ 1,...,n\right\} $, $j=\left\{ 1,...,m\right\}$ with $i=1,n$ or $j=1,m$ corresponding to the boundary $\Gamma$. Once again we substitute the two indices $i,j$ with the single running index $a$ and, for the collocation points $\boldsymbol{\hat{\tau}}_{a}$ in parametric coordinates, we denote the corresponding physical maps in the reference configuration as $\boldsymbol{\tau}_{a}$.

In isogeometric collocation, all test functions are chosen as Dirac delta functions centered at the interior and at the boundary collocation points. Applying the sifting properties of Eq. \eqref{eq:sifting_property} to the weighted residual form \eqref{eq:weighted_residual} results in
\begin{subequations}
\begin{alignat}{2}
\left[\nabla \cdot \boldsymbol{P}^{(k)} +\boldsymbol{B}^{(k)}\right]\left(\boldsymbol{\tau}_{a}^{(k)}\right)&=\boldsymbol{0} && \qquad\boldsymbol{\tau}_{a}^{(k)}\subset \Omega^{(k)},\\
\left[\boldsymbol{P}^{(k)} \boldsymbol{N}^{(k)} -\boldsymbol{\bar{T}}^{(k)} \right]\left(\boldsymbol{\tau}_{a}^{(k)}\right)&=\boldsymbol{0}&&\qquad\boldsymbol{\tau}_{a}^{(k)}\subset\mathrm{edge}^{(k)} \subset\Gamma_{N}^{(k)}, \\
\left[\boldsymbol{P}^{(k)}  \left( \boldsymbol{N}^{'(k)} + \boldsymbol{N}^{''(k)}  \right) - \left(\boldsymbol{\bar{T}}^{'(k)} +\boldsymbol{\bar{T}}^{''(k)} \right) \right]\left(\boldsymbol{\tau}_{a}^{(k)}\right)&=\boldsymbol{0}&&\qquad\boldsymbol{\tau}_{a}^{(k)}\equiv\mathrm{corner}^{(k)} \subset\Gamma_{N}^{(k)} ,\\
\left[\boldsymbol{P}^{(k)}\boldsymbol{N}^{(k)}-\boldsymbol{T}^{(k)}\right]\left(\boldsymbol{\tau}_{a}^{(k)}\right) &=\boldsymbol{0} &&\qquad\boldsymbol{\tau}_{a}^{(k)}\subset\mathrm{edge^{(k)}}\subset\Gamma_{C}^{(k)},\\
\left[\boldsymbol{P}^{(k)} \left( \boldsymbol{N}^{'(k)}+ \boldsymbol{N}^{''(k)} \right) - \left(\boldsymbol{T}^{'(k)}+\boldsymbol{T}^{''(k)}\right) \right]\left(\boldsymbol{\tau}_{a}^{(k)}\right)&=\boldsymbol{0}&&\qquad\boldsymbol{\tau}_{a}^{(k)}\equiv\mathrm{corner^{(k)}}\subset\Gamma_{C}^{(k)}.
\end{alignat}
\label{eq:StrongFormColHyperelasticity}
\end{subequations}
where the symbols $(\bullet)'$ and $(\bullet)''$ refer to the two adjacent edges meeting at a corner point. The different treatment for collocation points on edges and at corners is taken from \cite{auricchio2010isogeometric,auricchio2012isogeometric,de2015isogeometric,kruse2015isogeometric}. 

Thus, at the interior collocation points we obtain the strong form of the governing equations in the interior of the domain, whereas we recover the strong form of the Neumann boundary conditions and of the contact conditions at the collocation points located at the Neumann and at the contact boundary, respectively \cite{kruse2015isogeometric}. The Dirichlet boundary conditions are enforced strongly. 

As shown in \cite{de2015isogeometric}, the strong imposition of Neumann boundary conditions may lead to oscillations and thus to a loss of accuracy, in particular when non-uniform meshes are used. One possible remedy was introduced in \cite{de2015isogeometric} with the so-called enhanced collocation (EC) approach. The idea is to consider a combination of area and edge terms for the Neumann boundary conditions as follows
\begin{subequations}
\begin{alignat}{2}
&\left[\nabla \cdot \boldsymbol{P}^{(k)}+\boldsymbol{B}^{(k)}\right]\left(\boldsymbol{\tau}_{a}^{(k)}\right)&& \qquad\boldsymbol{\tau}_{a}^{(k)}\subset\mathrm{edge}^{(k)}\subset\Gamma_{N}^{(k)}, \\
&-\frac{C^{*}}{h^{(k)}}\left[\boldsymbol{P}^{(k)}\boldsymbol{N}^{(k)}-\boldsymbol{\bar{T}}^{(k)}\right]\left(\boldsymbol{\tau}_{a}^{(k)}\right)=\boldsymbol{0} \nonumber \\
&\left[\nabla \cdot \boldsymbol{P}^{(k)}+\boldsymbol{B}^{(k)}\right]\left(\boldsymbol{\tau}_{a}^{(k)}\right) && \qquad\boldsymbol{\tau}_{a}^{(k)}\equiv\mathrm{corner}^{(k)}\subset\Gamma_{N}^{(k)},\\
& -\frac{C^{*}}{h^{'(k)}}\left[\boldsymbol{P}^{(k)} \boldsymbol{N}^{'(k)} - \boldsymbol{\bar{T}}^{'(k)} \right]\left(\boldsymbol{\tau}_{a}^{(k)}\right) \nonumber \\
& -\frac{C^{*}}{h^{''(k)}}\left[\boldsymbol{P}^{(k)} \boldsymbol{N}^{''(k)} - \boldsymbol{\bar{T}}^{''(k)} \right]\left(\boldsymbol{\tau}_{a}^{(k)}\right) = \boldsymbol{0}  \nonumber
\end{alignat}
\label{eq:mod_coll_hyperelasticity}
\end{subequations}
where $h$ is the mesh size in the direction perpendicular to the edge. This approach requires a suitable choice for the constant $C^{*}$ in Eq.~\eqref{eq:mod_coll_hyperelasticity}. In \cite{de2015isogeometric}, $C^{*}$ was calibrated through numerical experiments and an optimal value of $C^{*}=4$ was found, which will also be used here. The EC approach is analogously applicable to the contact boundary, as already tested in \cite{kruse2015isogeometric}.

\section{Isogeometric collocated contact surface approach}
\label{sec:CollocatedContactSurfaceApproach}

\begin{figure}[htb]
\centering
\includegraphics[width=0.9\textwidth]{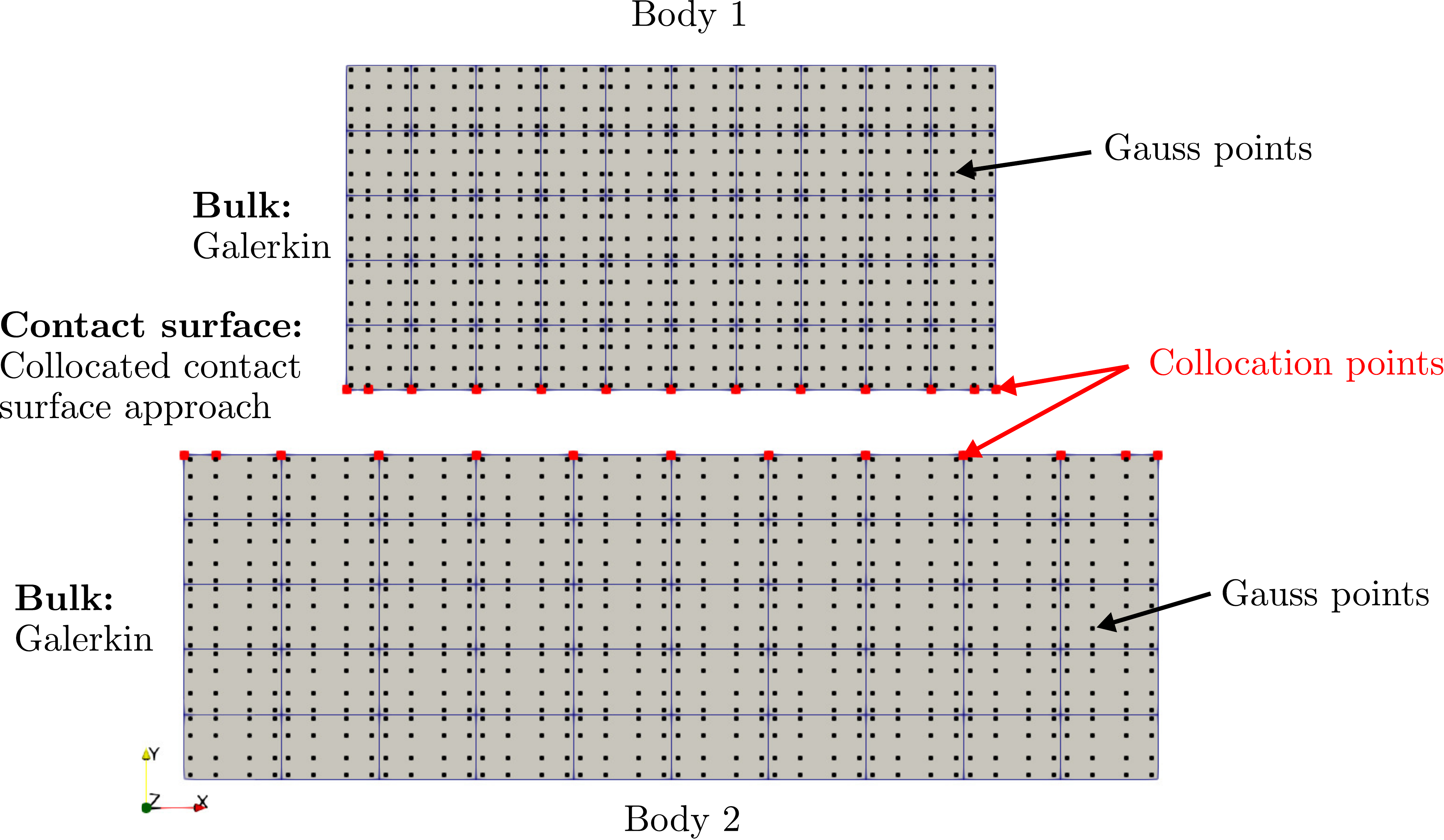}
\caption{Schematic representation of the isogeometric CCS approach. Discretization of each body with $10 \times 5$ elements of polynomial degree $p=3$.}
\label{fig:CollocatedContactSurfaceApproach}
\end{figure}

The idea of the CCS approach is to combine a contact formulation based on isogeometric collocation with a Galerkin treatment of the bulk, as sketched in Figure \ref{fig:CollocatedContactSurfaceApproach}. From the sketch it can be inferred that the proposed approach leads to a number of contact evaluation points significantly reduced compared to an integral contact formulation based e.g. on a standard Gaussian quadrature rule (like the GPTS approach), and equal to the number of evaluation points of pointwise contact formulations (like the PTS approach). 

The hybrid discretization of the CCS approach is obtained by adopting as test functions: 
\begin{itemize}
\item NURBS functions for the degrees of freedom corresponding to control points at the interior of the domain or at the Neumann boundaries;
\item Dirac delta functions centered at the appropriate collocation points for the degrees of freedom corresponding to control points at the contact boundaries.
\end{itemize}
The resulting expression of the test functions reads
\begin{equation}
\delta\mathbf{u}^{(k)}\left(\mathbf{X}\right)\approx\sum_{a=1}^{g^{(k)}}R_{a}\left(\mathbf{X}\right)\delta\hat{\mathbf{u}}_{a}^{(k)}+\sum_{a=g^{(k)}+1}^{N^{(k)}}\delta^D\left(\mathbf{X}-\boldsymbol{\tau}_{a}\right)\delta\hat{\mathbf{u}}_{a}^{(k)}
\end{equation}
For notational simplicity and without loss of generality, we
renumbered the control point variables in such a way that the first
$g^{(k)}$ are related to the interior and the Neumann boundary, whereas
the last $N^{(k)}-g^{(k)}$ are related to the contact boundary. Substitution in the weighted residual formulation \eqref{eq:weighted_residual} yields
\begin{eqnarray}
\sum_{k=1}^{2}\left\{ \sum_{a=1}^{g^{(k)}}\left[\delta\hat{\mathbf{u}}_{a}^{(k)}\cdot\int_{\Omega^{(k)}}\left(\nabla\cdot\boldsymbol{P}^{(k)}+\mathbf{B}^{(k)}\right)R_{a}d\varOmega\right.\right.\nonumber \\
-\delta\hat{\mathbf{u}}_{a}^{(k)}\cdot\int_{\Gamma_{N}^{(k)}}\left(\boldsymbol{P}^{(k)}\boldsymbol{N}^{(k)}-\mathbf{\bar{T}}^{(k)}\right)R_{a}d\Gamma\nonumber \\
\left.-\delta\hat{\mathbf{u}}_{a}^{(k)}\cdot\int_{\Gamma_{C}^{(k)}}\left(\boldsymbol{P}^{(k)}\boldsymbol{N}^{(k)}-\mathbf{T}^{(k)}\right)R_{a}d\Gamma\right]\nonumber \\
+\sum_{a=g^{(k)}+1}^{N^{(k)}}\left[\delta\hat{\mathbf{u}}_{a}^{(k)}\cdot\int_{\Omega^{(k)}}\left(\nabla\cdot\boldsymbol{P}^{(k)}+\mathbf{B}^{(k)}\right)\delta^D\left(\mathbf{X}-\boldsymbol{\tau}_{a}\right)d\varOmega\right.\nonumber \\
-\delta\hat{\mathbf{u}}_{a}^{(k)}\cdot\int_{\Gamma_{N}^{(k)}}\left(\boldsymbol{P}^{(k)}\boldsymbol{N}^{(k)}-\mathbf{\bar{T}}^{(k)}\right)\delta^D\left(\mathbf{X}-\boldsymbol{\tau}_{a}\right)d\Gamma\nonumber \\
\left.\left.-\delta\hat{\mathbf{u}}_{a}^{(k)}\cdot\int_{\Gamma_{C}^{(k)}}\left(\boldsymbol{P}^{(k)}\boldsymbol{N}^{(k)}-\mathbf{T}^{(k)}\right)\delta^D\left(\mathbf{X}-\boldsymbol{\tau}_{a}\right)d\Gamma\right]\right\}  & = & 0
\end{eqnarray}
Since $R_{a}$ for $a=1,...,g^{(k)}+1$ vanish on $\Gamma_{C}^{(k)}$
and since $\boldsymbol{\tau}_{a}$ for $a=g^{(k)}+1,...,N^{(k)}$
are located on $\Gamma_{C}^{(k)}$, the above discretized weighted
residual form reduces to
\begin{eqnarray}
\sum_{k=1}^{2}\left\{ \sum_{a=1}^{g^{(k)}}\left[\delta\hat{\mathbf{u}}_{a}^{(k)}\cdot\int_{\Omega^{(k)}}\left(\nabla\cdot\boldsymbol{P}^{(k)}+\mathbf{B}^{(k)}\right)R_{a}d\varOmega\right.\right.\nonumber \\
\left.-\delta\hat{\mathbf{u}}_{a}^{(k)}\cdot\int_{\Gamma_{N}^{(k)}}\left(\boldsymbol{P}^{(k)}\boldsymbol{N}^{(k)}-\mathbf{\bar{T}}^{(k)}\right)R_{a}d\Gamma\right]\nonumber \\
\left.-\sum_{a=g^{(k)}+1}^{N^{(k)}}\delta\hat{\mathbf{u}}_{a}^{(k)}\cdot\int_{\Gamma_{C}^{(k)}}\left(\boldsymbol{P}^{(k)}\boldsymbol{N}^{(k)}-\mathbf{T}^{(k)}\right)\delta^D\left(\mathbf{X}-\boldsymbol{\tau}_{a}\right)d\Gamma\right\}  & = & 0\label{eq:CCS}
\end{eqnarray}
In Eq. (\ref{eq:CCS}), the integrals in the first row can be integrated
by parts, delivering the ``usual'' Galerkin contributions to the
residual vector (and, upon linearization, to the tangent stiffness
matrix), whereas the integral in the second row, due to the sifting
property in Eq. \eqref{eq:sifting_property}$_2$, delivers the collocated contact contributions.
With this approach, which is reminiscent of (but different from) the hybrid collocation-Galerkin treatment in \cite{de2015isogeometric}, a Galerkin formulation for the interior and the Neumann boundaries and a collocated formulation for the contact boundaries are naturally obtained. It was shown in \cite{de2015isogeometric,kruse2015isogeometric} that in the framework of isogeometric collocation a simple pointwise contact treatment combined with the two-half-pass algorithm and the penalty method passes the patch test to machine precision and delivers accurate results. Hence, the CCS approach is expected to inherit these performance features, while keeping the flexibility and accuracy of Galerkin for the bulk behavior.

\begin{figure}[htb]
\centering
\includegraphics[width=0.4\textwidth]{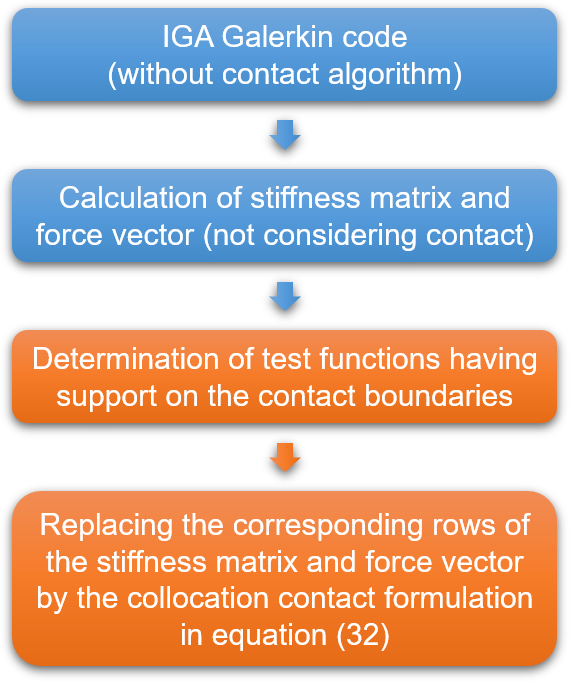}
\caption{Schematic representation of the implementation procedure of the CCS approach.}
\label{fig:FlowchartCollocatedContactAlgorithm}
\end{figure}

The implementation strategy to endow a standard IGA Galerkin formulation with the CCS approach is illustrated in Figure \ref{fig:FlowchartCollocatedContactAlgorithm}. From an operational standpoint, the incorporation into a pre-existing IGA Galerkin code is straightforward. This code is first used to calculate the global stiffness matrix and residual vector for both bodies with the standard Galerkin formulation, not taking into account the contact boundaries. Afterwards all test functions having support on the contact boundaries $\Gamma_C^{(k)}$ and their global indices have to be identified. Subsequently the rows of the stiffness matrix and residual vector corresponding to these indices are completely substituted by the collocation based contact contributions computed as in \eqref{eq:StrongFormColHyperelasticity}d,e. This substitution is easily carried out directly in the final system of linear equations, with no need for manipulations at the element level of the Galerkin code. 

It is evident that the incorporation of the collocation contact formulation is very similar and equivalently simple as the treatment of Dirichlet boundary conditions. For frictionless contact, a drawback of the approach is the loss of symmetry of the tangent stiffness matrix. However, this is no longer an issue in the more realistic situation of frictional contact, in which the tangent stiffness matrix is asymmetric in all cases. In case of inactive contact, the collocation-based contact formulation automatically enforces homogeneous Neumann boundary conditions, hence there is no need for segmentation of the contact surfaces. 

\begin{figure}[htb]
\centering
\includegraphics[width=0.78\textwidth]{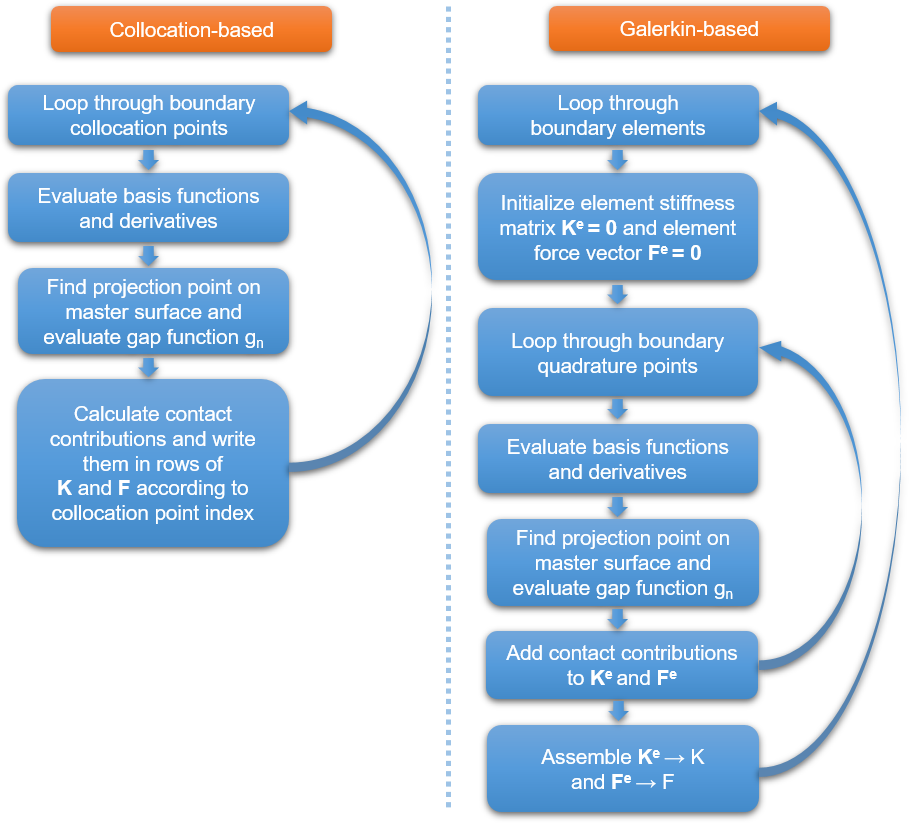}
\caption{Flowchart: Comparison of collocation-based and Galerkin-based penalty contact approaches.}
\label{fig:FlowchartContactCode}
\end{figure}

A flowchart comparing the general implementation of collocation-based and Galerkin-based penalty contact approaches within the framework of an IGA simulation can be found in Figure \ref{fig:FlowchartContactCode}.

\section{Numerical examples}
\label{sec:NumericalExamples}

In the previous sections, several contact approaches were introduced. For ease of reference, their main features are summarized in Table \ref{tab:MethodsAbbreviations}, where the proposed CCS approach is also included.  In this section we consider four different numerical examples to investigate the performance of the proposed CCS approach in comparison with that of the other approaches.  We consider here two-dimensional problems under plane strain conditions. 

\begin{table}[htb]
\centering
\begin{tabular}{L{2.1cm}|L{2.7cm}|L{2cm}|L{6cm}}
\toprule                                                                                                          
Type & Name & Abbr. & Features\\
\midrule                                                                                                         
\multirow{2}{\linewidth}{Galerkin-based}    & Gauss-point-to-segment / with two-half-pass    & GPTS /  GPTS-2hp     & Weak contact enforcement at Gauss quadrature points on slave surface / on both contact surfaces         \\ \cline{2-4} 
                                   & Point-to segment / with two-half-pass          & PTS / PTS-2hp     & Weak contact enforcement at collocation points on slave surface / on both contact surfaces                  \\ \hline
\multirow{2}{\linewidth}{Collocation-based}    & Collocation / Enhanced collocation     & C / EC  & Strong /enhanced contact enforcement at collocation points on both contact surfaces, collocation treatment of the bulk        \\ \cline{2-4} 
                                   & Collocated contact surface / Enhanced collocated contact surface       & CCS / ECCS     & Strong /enhanced contact enforcement at collocation points on both contact surfaces, Galerkin treatment of the bulk \\ \hline
\end{tabular} 
\caption{Overview of the considered contact algorithms.}                                                                                                                                                                               
\label{tab:MethodsAbbreviations}                            
\end{table}

\subsection{Contact patch test}
\label{subsec:ContactPatchTest}

\begin{figure}[H]
\centering
\begin{subfigure}[b]{0.46\textwidth}
\includegraphics[width=\textwidth]{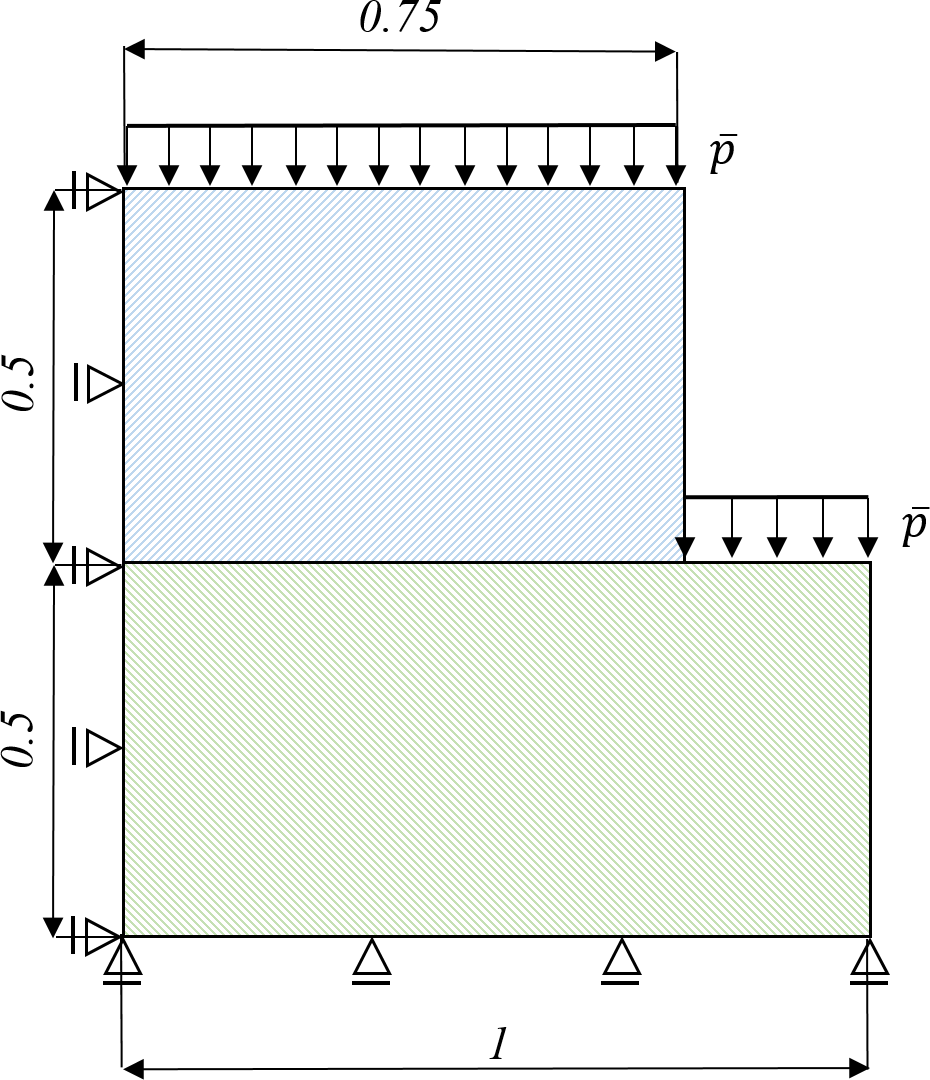}
\caption{Geometry and boundary conditions}
\end{subfigure}
\qquad
\begin{subfigure}[b]{0.42\textwidth}
\begin{mdframed}[frametitle={\footnotesize Simulation setup},userdefinedwidth=\textwidth]
\begin{footnotesize}
Number of control points: $n_{cp} = 30\times30$\\
Polynomial degree: $p = 2$\\
Penalty parameter: $\epsilon_n = 100$\\
$1^{st}$ Lamé param.: $\mu=0.5$\\
$2^{nd}$ Lamé param.: $\lambda=1$\\
Number of loadsteps: $n_l= 10$\\
Applied pressure: $\bar{p} = 0.01$
\end{footnotesize}
\end{mdframed}\vspace{60pt}
\caption{Simulation parameters}
\end{subfigure}
\caption{Contact patch test: Geometry, boundary conditions and simulation setup.}
\label{fig:GeometryPatchTest}
\end{figure}

The first numerical example consists of the so-called contact patch test, proposed by Taylor and Papadopoulos in \cite{taylor1991patch}. The main objective of this setup is to test the capability of a contact formulation to transfer a constant contact pressure across the interface between two bodies discretized with non-conforming meshes. The geometry, boundary conditions and simulation parameters are depicted in Figure \ref{fig:GeometryPatchTest}. The two blocks are pressed onto each other  with a uniform pressure $\bar{p} = 0.01$, which is applied within ten loadsteps. Symmetry boundary conditions are applied on the left vertical edges of both blocks. The bottom boundary of the lower block is fixed in vertical direction and homogeneous Neumann boundary conditions are applied in horizontal direction. Since the considered deformations are comparatively small, linear elasticity is assumed for this example. 

\begin{figure}[H]
\centering
\begin{subfigure}[b]{0.36\textwidth}
\includegraphics[width=\textwidth]{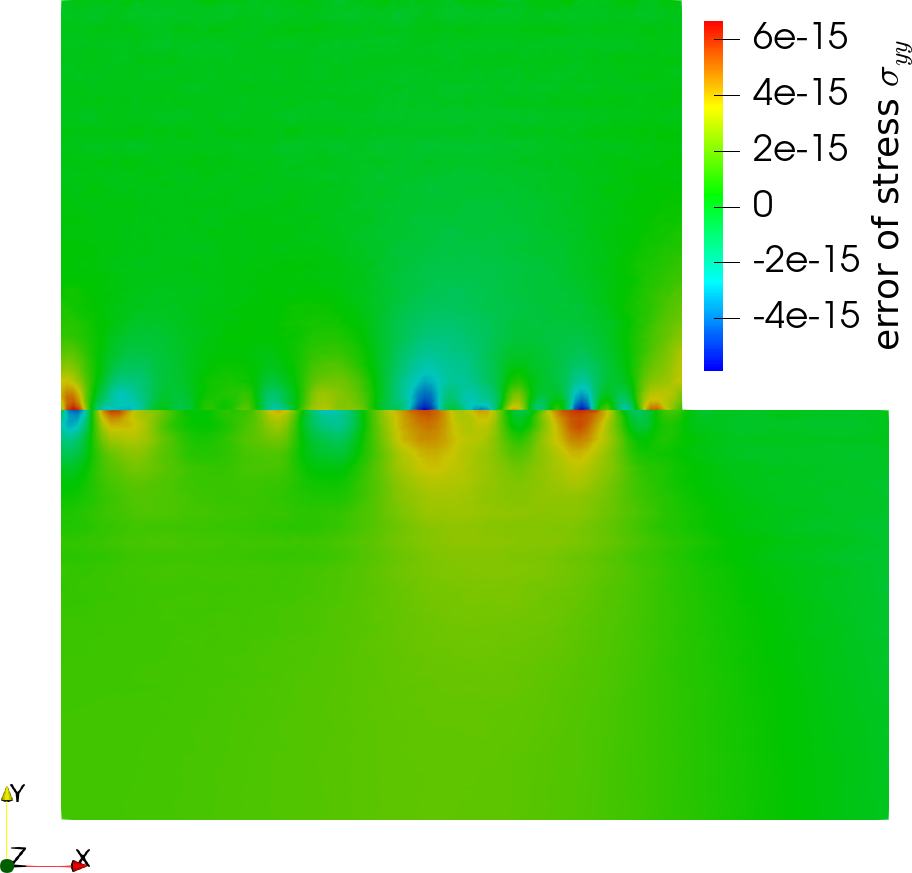}
\caption{CCS}
\end{subfigure}
\qquad
\begin{subfigure}[b]{0.36\textwidth}
\includegraphics[width=\textwidth]{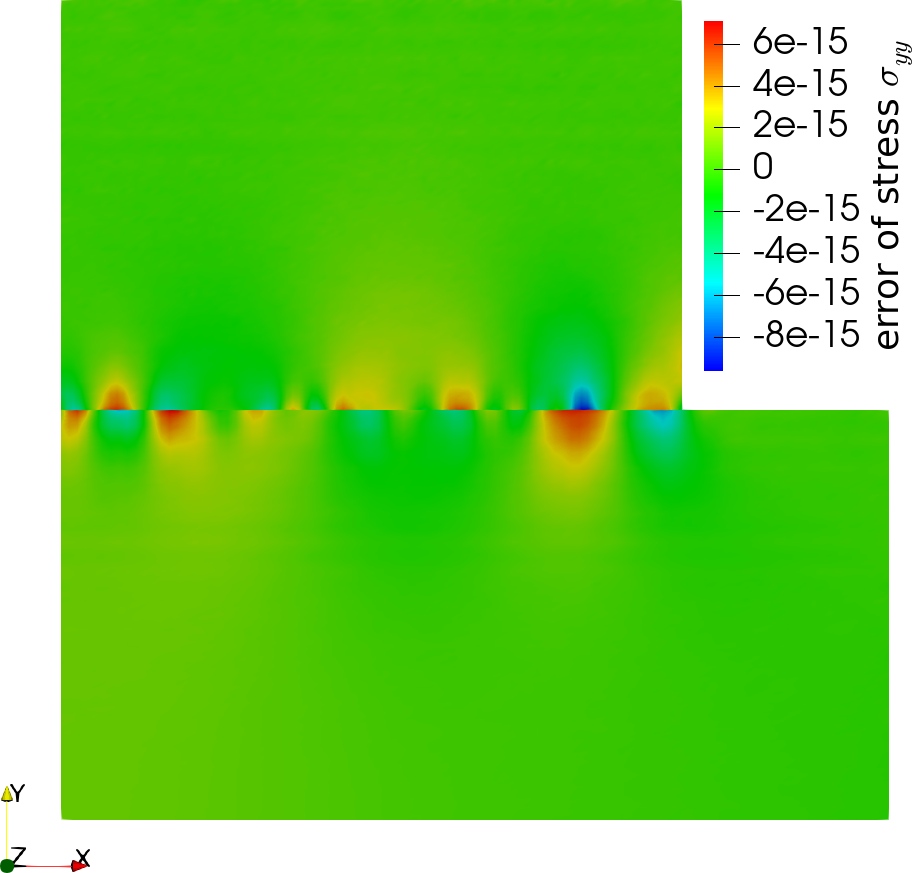}
\caption{ECCS}
\end{subfigure}\vspace{6pt}\\
\begin{subfigure}[b]{0.36\textwidth}
\includegraphics[width=\textwidth]{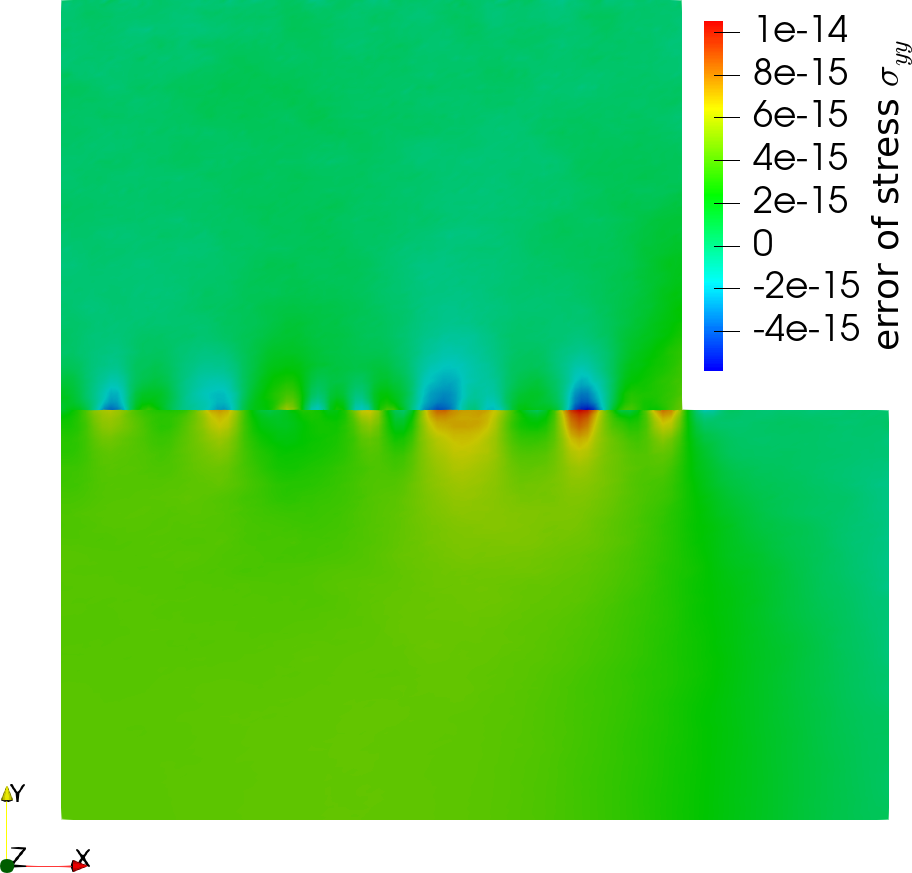}
\caption{C}
\end{subfigure}
\qquad
\begin{subfigure}[b]{0.36\textwidth}
\includegraphics[width=\textwidth]{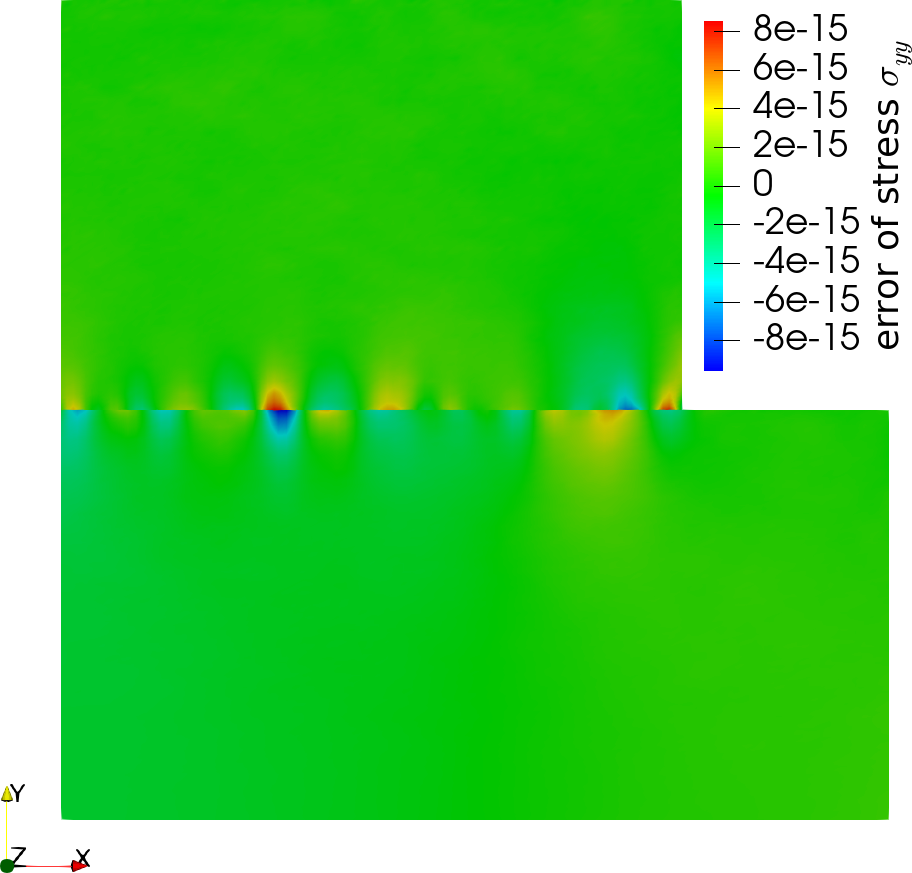}
\caption{EC}
\end{subfigure}
\caption{Contact patch test: Error of stress $\sigma_{yy}$ for the proposed collocated contact surface approaches (CCS \& ECCS), collocation (C) and enhanced collocation (EC).}
\label{fig:ErrorPatchTestCollocation}
\end{figure}

\begin{figure}[H]
\centering
\begin{subfigure}[b]{0.36\textwidth}
\includegraphics[width=\textwidth]{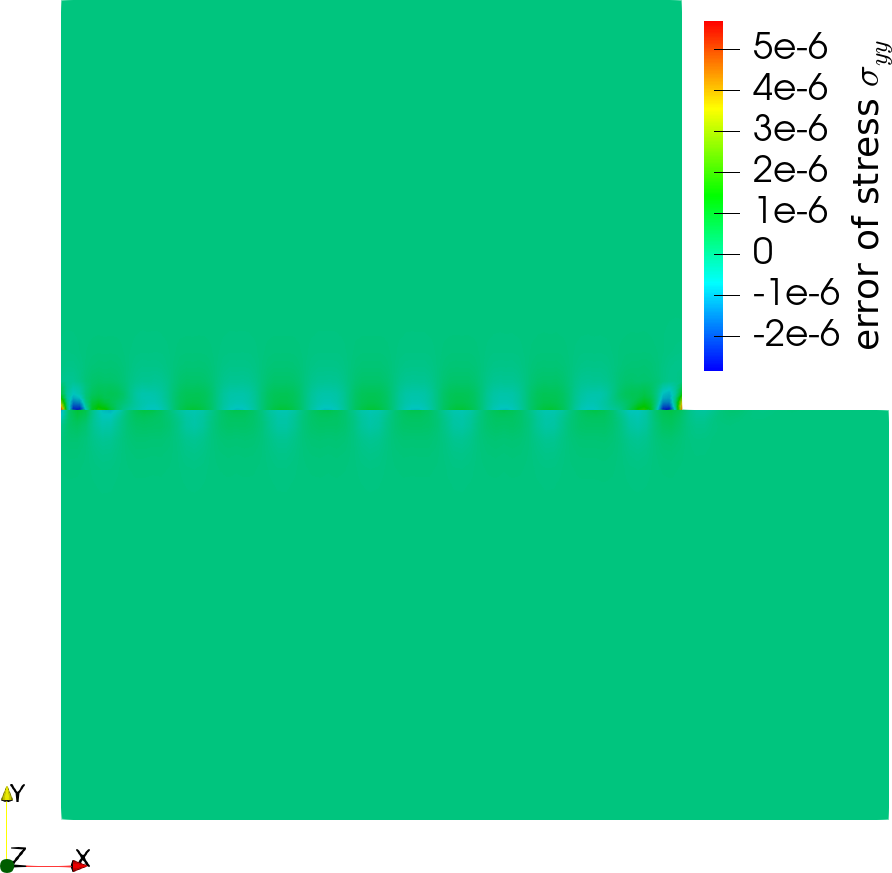}
\caption{GPTS}
\end{subfigure}
\qquad
\begin{subfigure}[b]{0.36\textwidth}
\includegraphics[width=\textwidth]{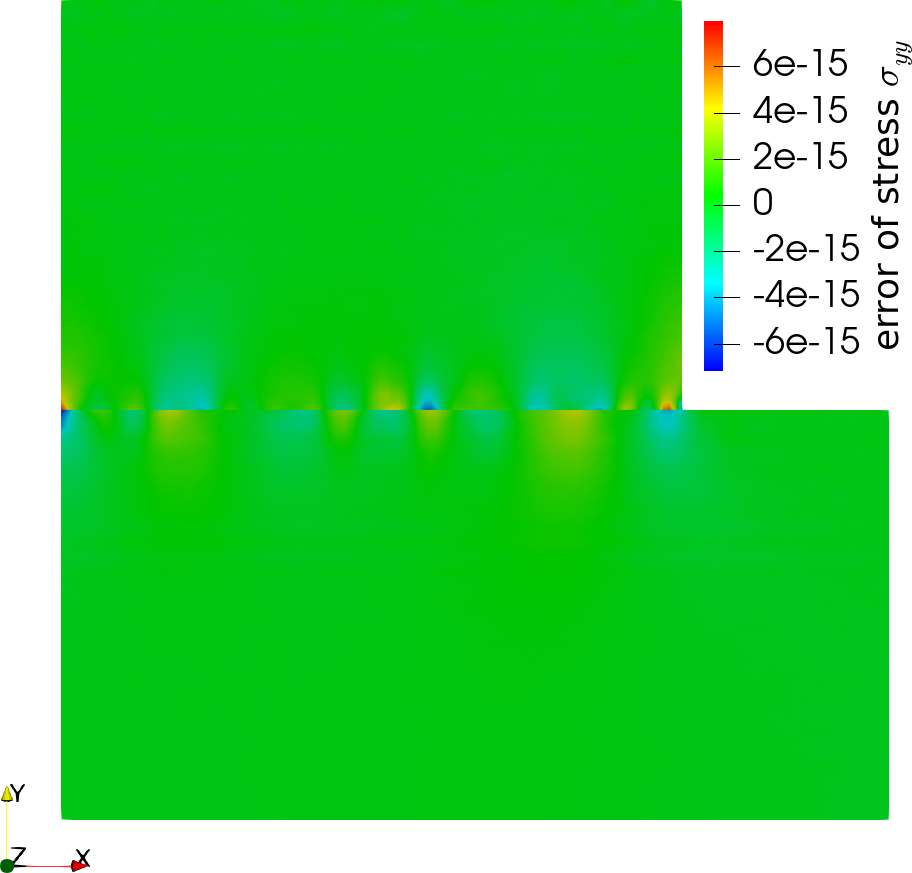}
\caption{GPTS-2hp}
\end{subfigure}\vspace{6pt}\\
\begin{subfigure}[b]{0.36\textwidth}
\includegraphics[width=\textwidth]{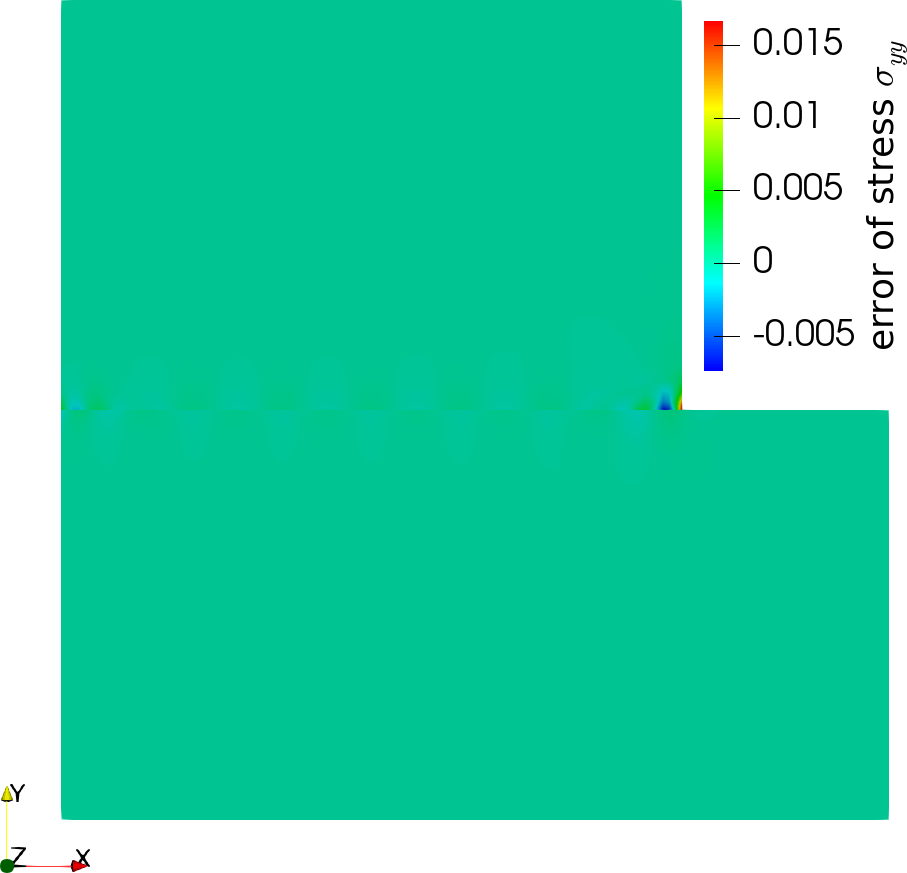}
\caption{PTS}
\end{subfigure}
\qquad
\begin{subfigure}[b]{0.36\textwidth}
\includegraphics[width=\textwidth]{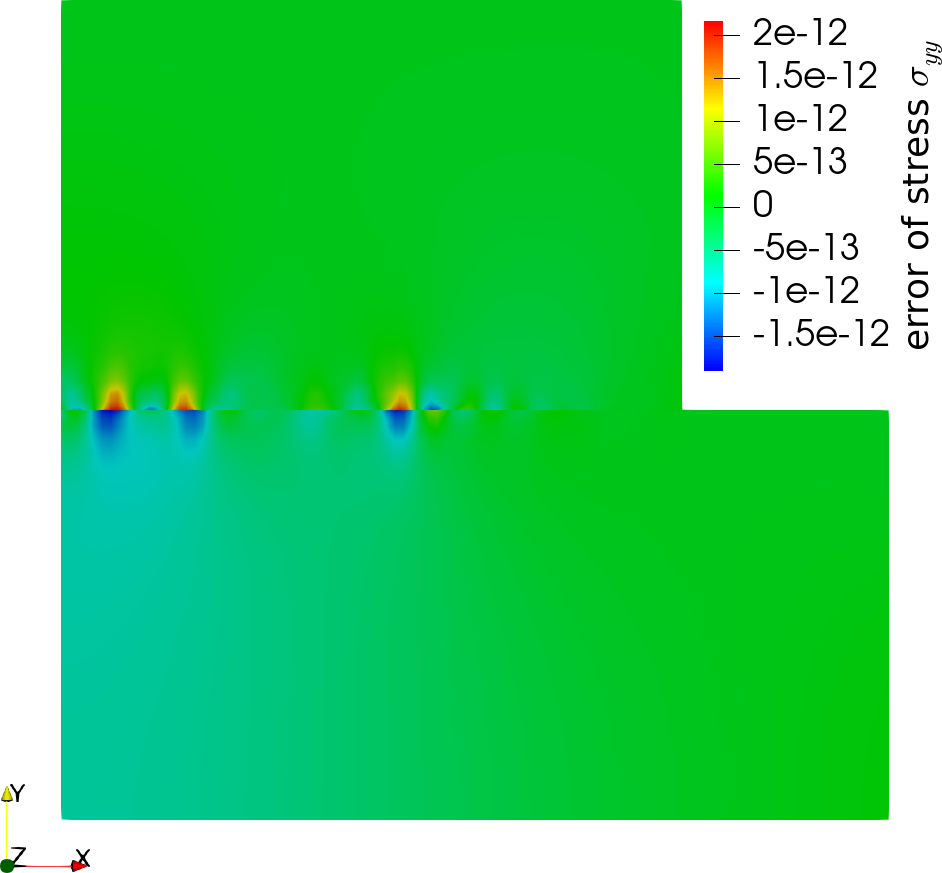}
\caption{PTS-2hp}
\end{subfigure}
\caption{Contact patch test: Error of stress $\sigma_{yy}$ for the Gauss-point-to-segment (GPTS) and Point-to-segment (PTS) approaches and the corresponding two-half-pass (2hp) formulations.}
\label{fig:ErrorPatchTestCompMethods}
\end{figure}

The resulting errors of the stress component $\sigma_{yy}$ are shown in Figures \ref{fig:ErrorPatchTestCollocation} and \ref{fig:ErrorPatchTestCompMethods}. Figure \ref{fig:ErrorPatchTestCollocation} contains the error plots for the newly proposed CCS and ECCS approaches. For comparison, results of the corresponding full collocation (C and EC) approaches are also shown. As expected, all four methods fulfil the contact patch test to machine precision, i.e. the collocated contact formulation in CCS and ECCS preserves the properties of the same formulation in a fully collocated context \cite{de2014isogeometric}. For this case featuring nearly homogeneous meshes, EC and ECCS perform nearly identically to C and CCS. 

Figure \ref{fig:ErrorPatchTestCompMethods} displays the error plots for the GPTS and PTS approaches. It is known from the literature (see e.g. \cite{de2014isogeometric}), that the GPTS approach is only able to fulfil the contact patch test up to the integration error. This is confirmed by the results in Figure \ref{fig:ErrorPatchTestCompMethods}. The extension to a two-half-pass formulation brings the error down to machine precision.
The PTS formulation leads to a higher error than with the GPTS approach. The two-half-pass extension significantly improves the results, but does not reach machine precision, as expected for a pointwise approach (despite the enhancement of weight computation in \cite{matzen2016weighted}).

\subsection{Two deformable blocks}
\label{subsec:TwoDeformableBlocks}

\begin{figure}[H]
\centering
\begin{subfigure}[b]{0.5\textwidth}
\includegraphics[width=\textwidth]{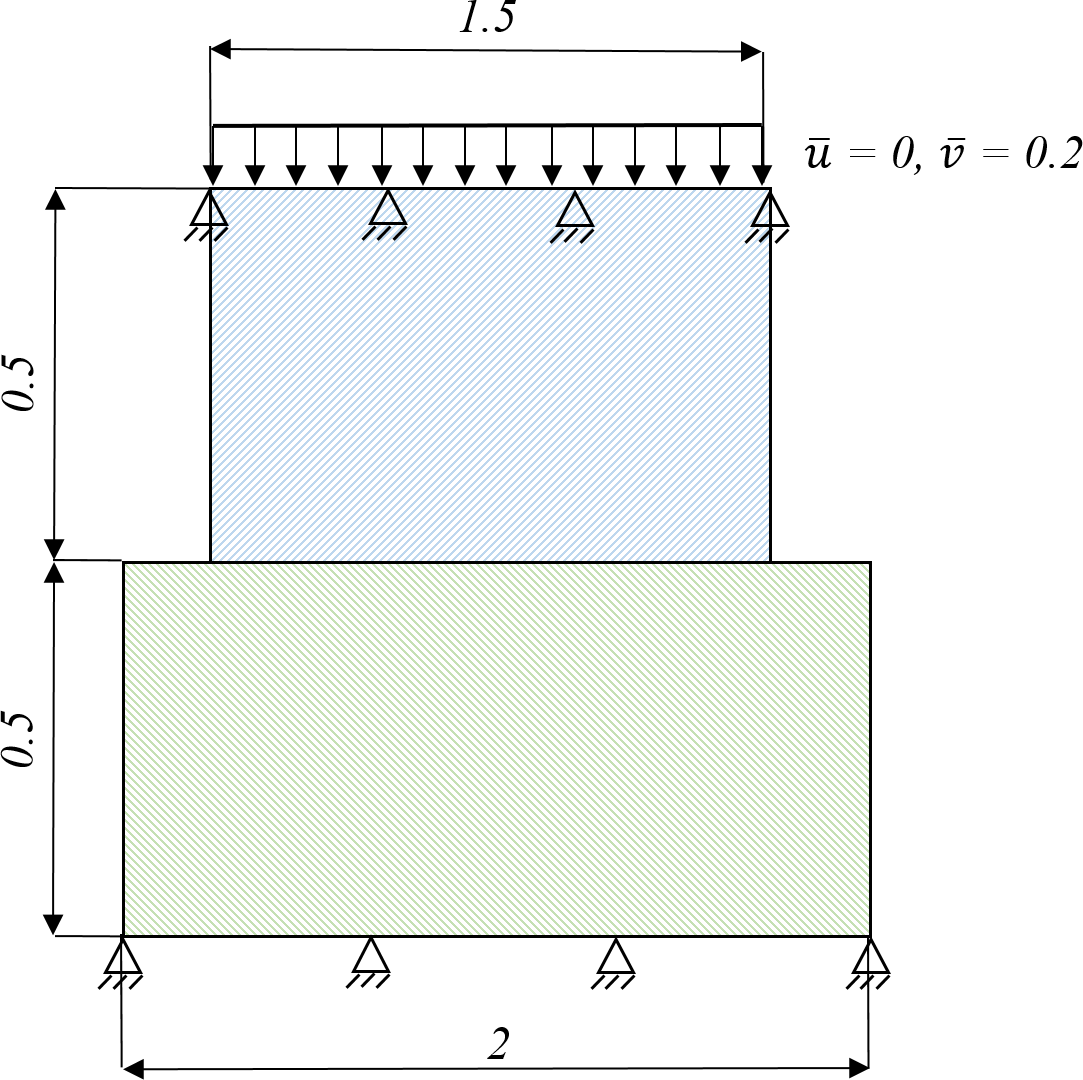}
\caption{Geometry and boundary conditions}
\end{subfigure}
\quad
\begin{subfigure}[b]{0.42\textwidth}
\begin{mdframed}[frametitle={\footnotesize Simulation setup},userdefinedwidth=\textwidth]
\begin{footnotesize}
Number of control points: \\$n_{cp} = 10\times15$ / $25\times10$\\
Polynomial degree: $p = 2$\\
Penalty parameter: $\epsilon_n = 1500$\\
$1^{st}$ Lamé param.: $\mu=0.5$\\
$2^{nd}$ Lamé param.: $\lambda=0.5$\\
Number of loadsteps: $n_l= 20$
\end{footnotesize}
\end{mdframed}\vspace{60pt}
\caption{Simulation parameters}
\end{subfigure}
\caption{Two deformable blocks: Geometry, boundary conditions and simulation setup.}
\label{fig:GeometryTwoDeformableBlocks}
\end{figure}

In the following numerical example, which was initially presented in \cite{de2015isogeometric}, two deformable blocks are pressed against each other. Geometry, boundary conditions and further simulation parameters are illustrated in Figure \ref{fig:GeometryTwoDeformableBlocks}. A uniform vertical displacement $\bar{v} = 0.2$ and zero horizontal displacement are enforced on the upper edge of the upper block. Although the assumption of small deformations is clearly violated in this example, we adopt a linearly elastic material model to adhere to the original simulation setup \cite{de2015isogeometric}.

Two different discretizations are tested in order to study the effect of the element aspect ratio on the results. The first discretization consists of $10\times15$ control points for each body. Hence, the element size is larger in the horizontal direction, i.e. in the direction perpendicular to the vertical edges of the blocks where homogeneous Neumann boundary conditions are applied. In \cite{de2015isogeometric} this element shape was found to lead to oscillations with the pure collocation scheme. 

\begin{figure}[H]
\centering
\begin{subfigure}[b]{0.48\textwidth}
\includegraphics[width=\textwidth]{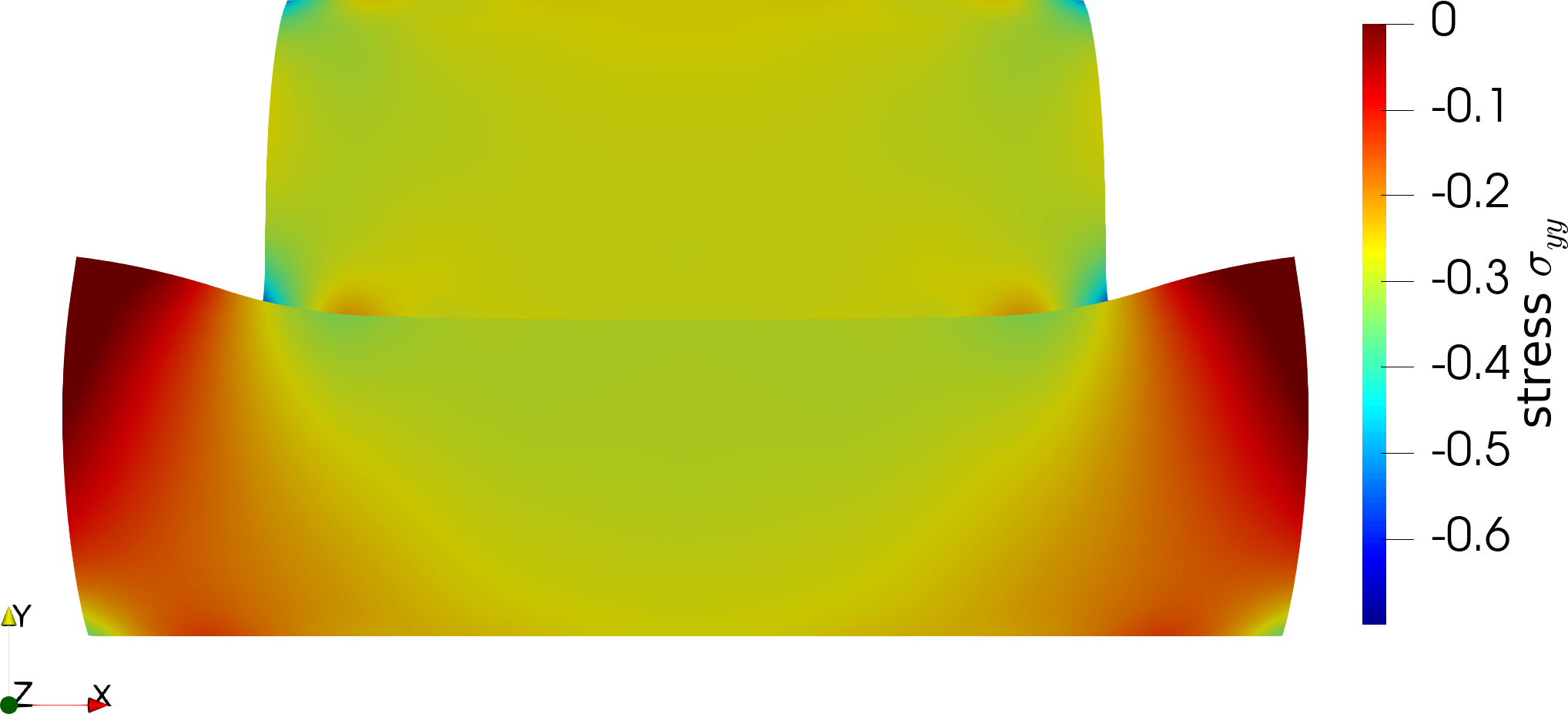}
\caption{CCS}
\end{subfigure}
\quad
\begin{subfigure}[b]{0.48\textwidth}
\includegraphics[width=\textwidth]{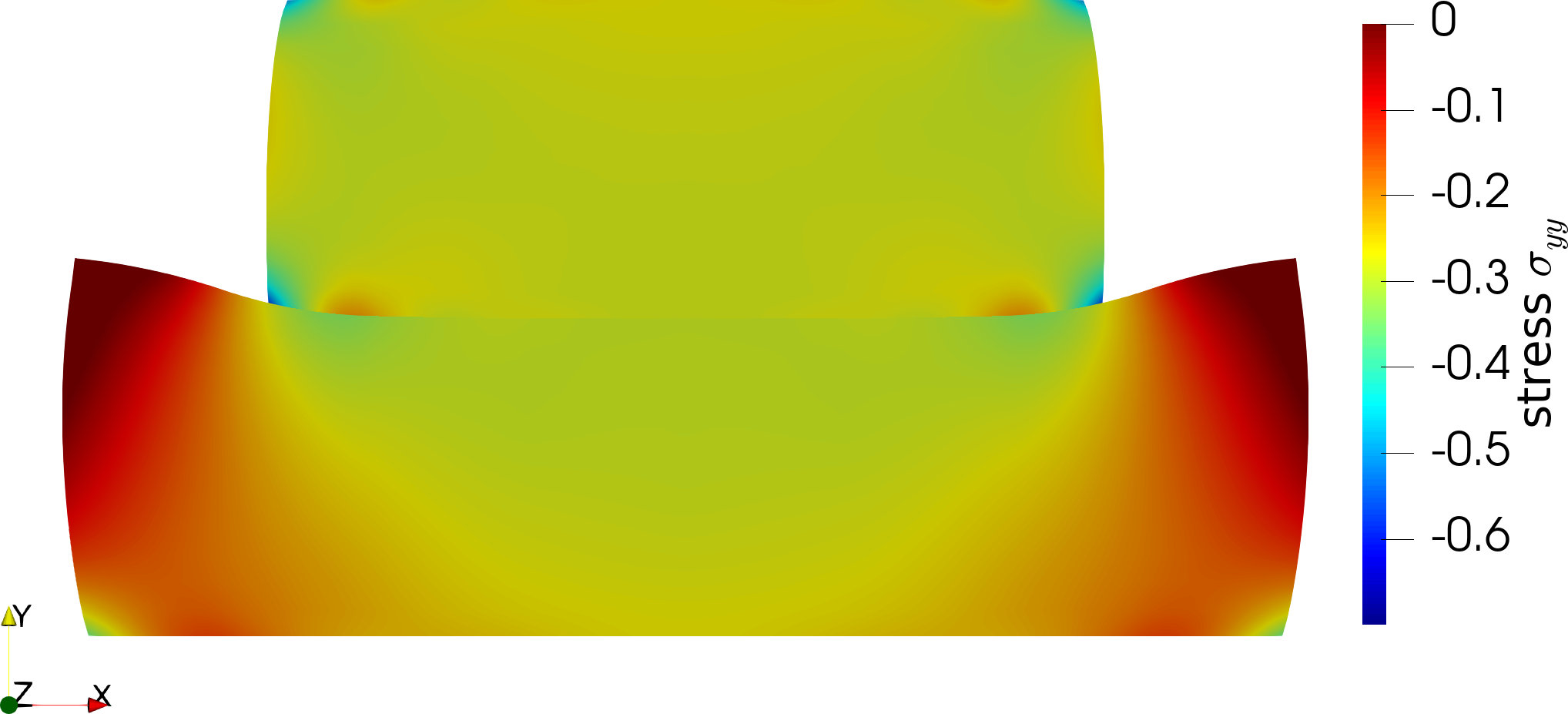}
\caption{ECCS}
\end{subfigure}\vspace{6pt}\\
\begin{subfigure}[b]{0.48\textwidth}
\includegraphics[width=\textwidth]{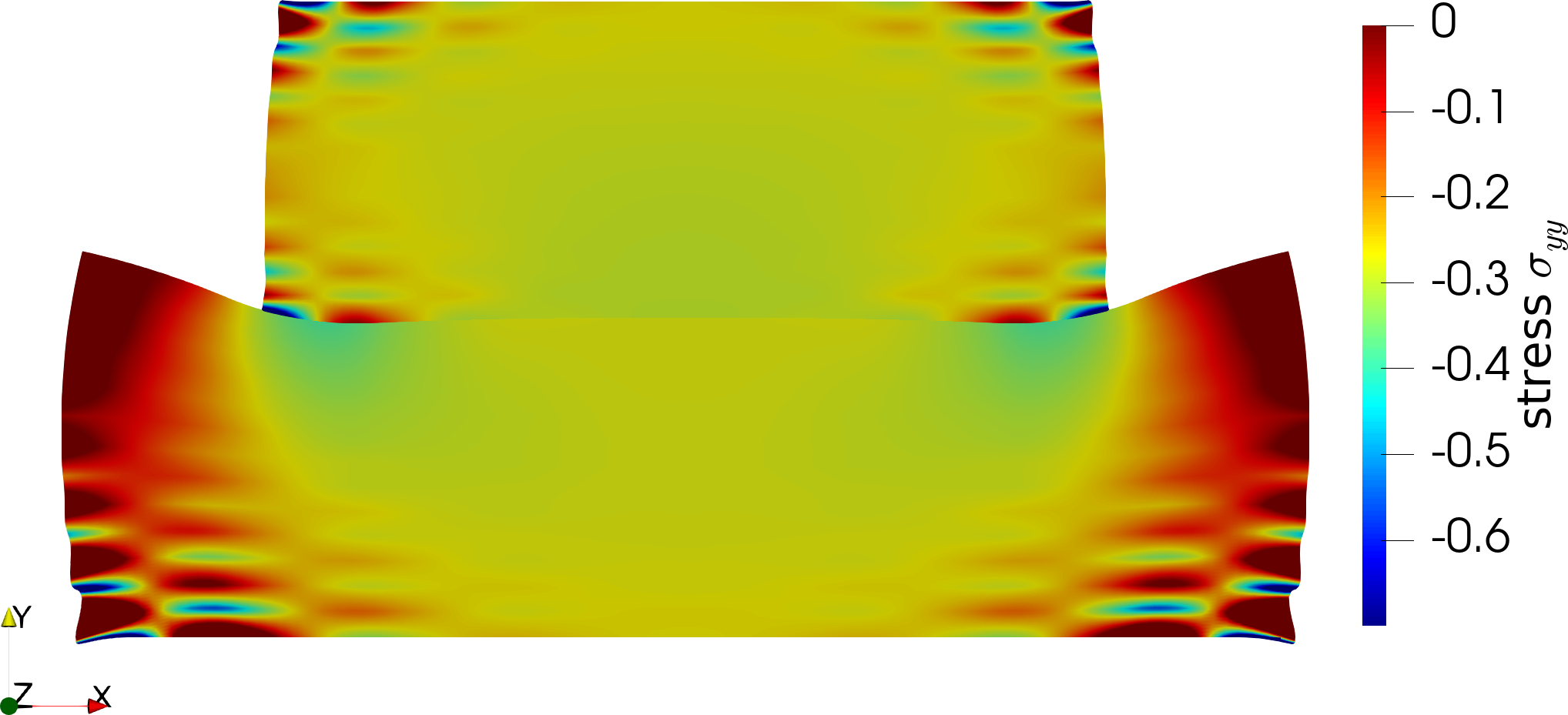}
\caption{C}
\end{subfigure}
\quad
\begin{subfigure}[b]{0.48\textwidth}
\includegraphics[width=\textwidth]{./images/solutionfields/BlockOnBlock/BlockOnBlockModcollGal_10_15_p2_Syy.png}
\caption{EC}
\end{subfigure}
\caption{Two deformable blocks: Stress $\sigma_{yy}$ for the proposed collocated contact surface approaches (CCS \& ECCS), collocation (C) and enhanced collocation (EC). Discretization of each body with $10 \times 15$ control points.}
\label{fig:StressTwoBlock1015Collocation}
\end{figure}

\begin{figure}[H]
\centering
\begin{subfigure}[b]{0.48\textwidth}
\includegraphics[width=\textwidth]{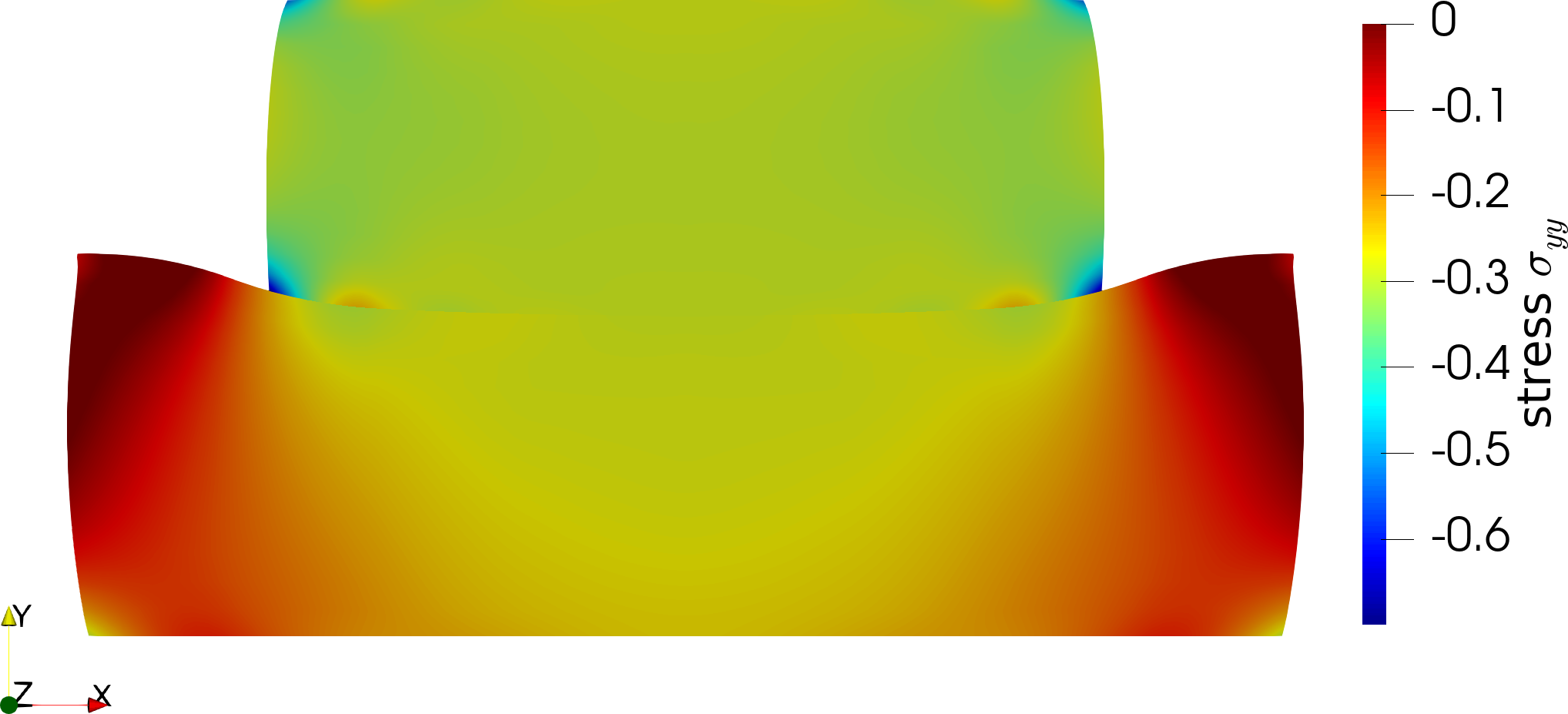}
\caption{GPTS}
\end{subfigure}
\quad
\begin{subfigure}[b]{0.48\textwidth}
\includegraphics[width=\textwidth]{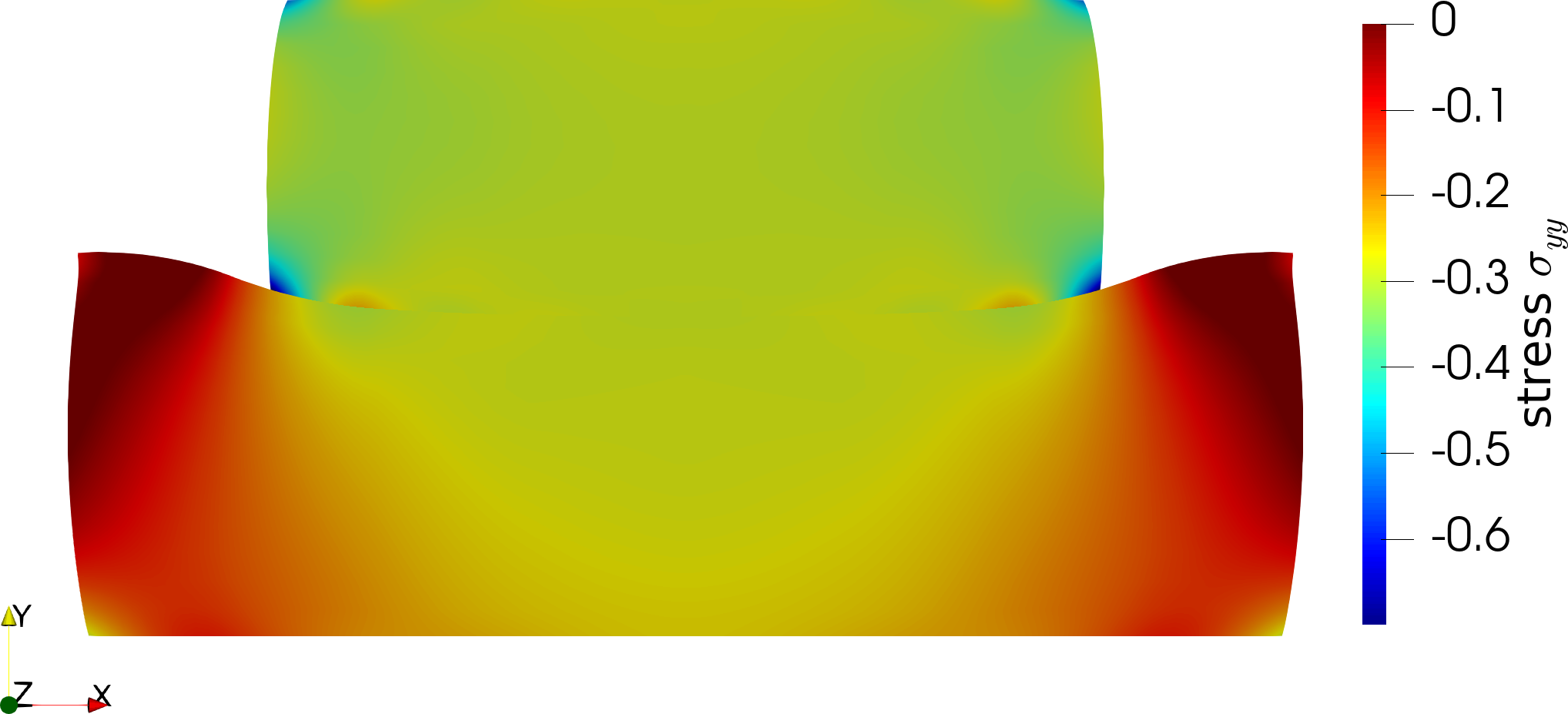}
\caption{GPTS-2hp}
\end{subfigure}\vspace{6pt}\\
\begin{subfigure}[b]{0.48\textwidth}
\includegraphics[width=\textwidth]{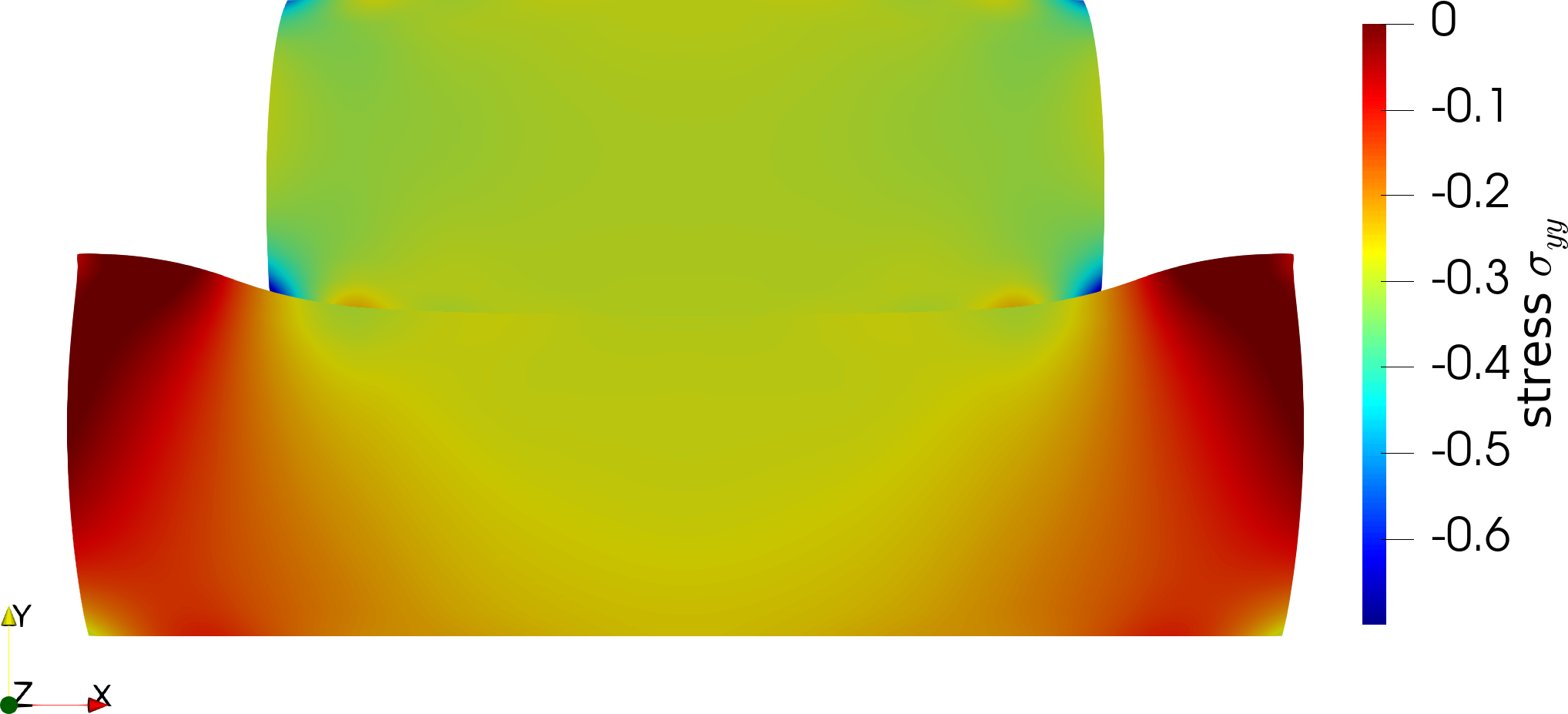}
\caption{PTS}
\end{subfigure}
\quad
\begin{subfigure}[b]{0.48\textwidth}
\includegraphics[width=\textwidth]{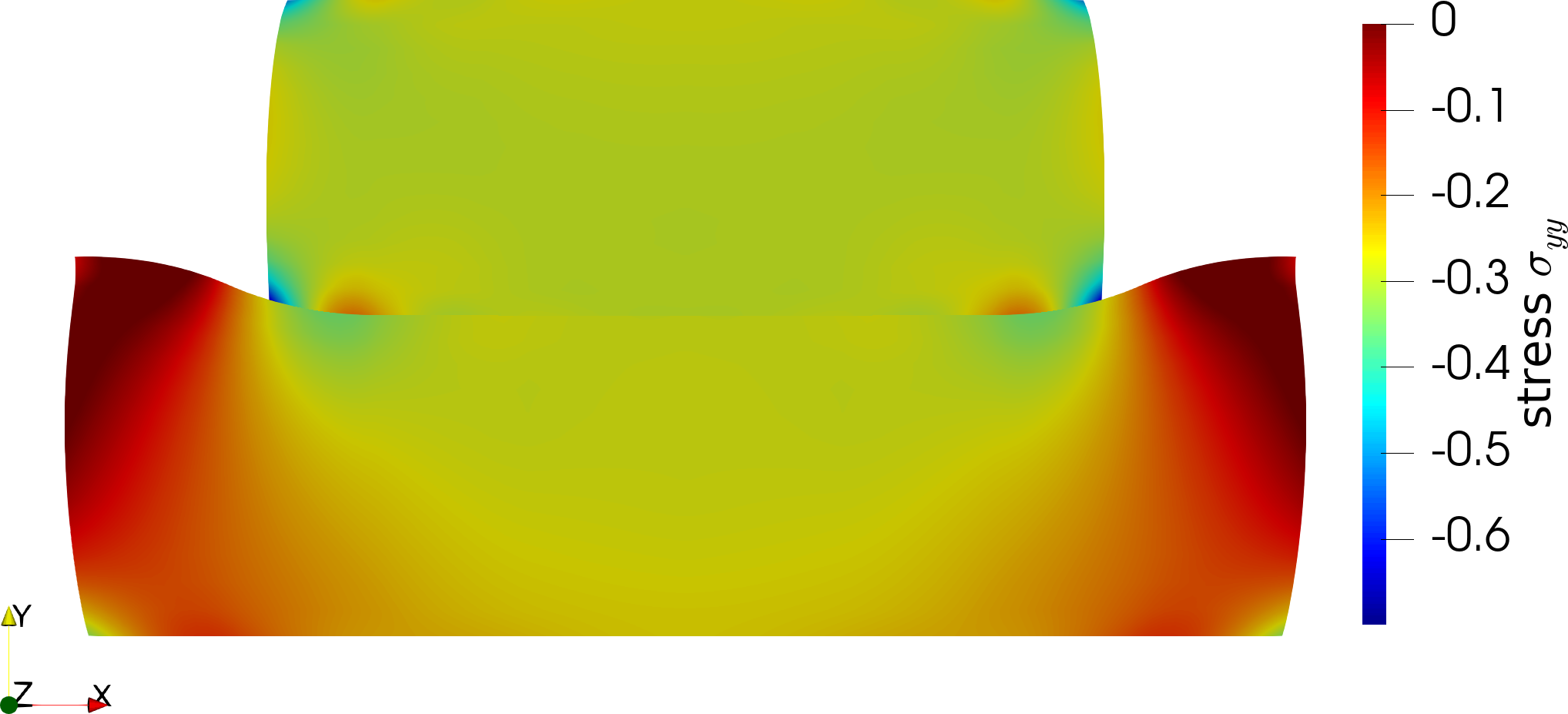}
\caption{PTS-2hp}
\end{subfigure}
\caption{Two deformable blocks: Stress $\sigma_{yy}$ for the Gauss-point-to-segment (GPTS) and Point-to-segment (PTS) approaches and the corresponding two-half-pass (2hp) formulations. Discretization of each body with $10 \times 15$ control points.}
\label{fig:StressTwoBlock1015CompMethods}
\end{figure}

The resulting plots of the stress component $\sigma_{yy}$ for this discretization are given in Figures \ref{fig:StressTwoBlock1015Collocation} and \ref{fig:StressTwoBlock1015CompMethods}. Figure \ref{fig:StressTwoBlock1015Collocation} contains the results for the C, EC and the proposed CCS and ECCS approaches. Strong oscillations appear with pure collocation (C). The EC approach successfully suppresses these oscillations, which it was intended for. Interestingly, the proposed CCS approach is also free of oscillations, probably due to its Galerkin treatment of the Neumann boundary conditions. Thus its enhancement as in ECCS - although performing well - is not required. 

The plots of the stress component $\sigma_{yy}$ for the GPTS and PTS approaches are given in Figure \ref{fig:StressTwoBlock1015CompMethods}. As expected, for these approaches no oscillations are obtained. The stress plots of all the different approaches (except for pure collocation) look similar, despite the coarse discretization. 

\begin{figure}[H]
\centering
\begin{subfigure}[b]{0.48\textwidth}
\includegraphics[width=\textwidth]{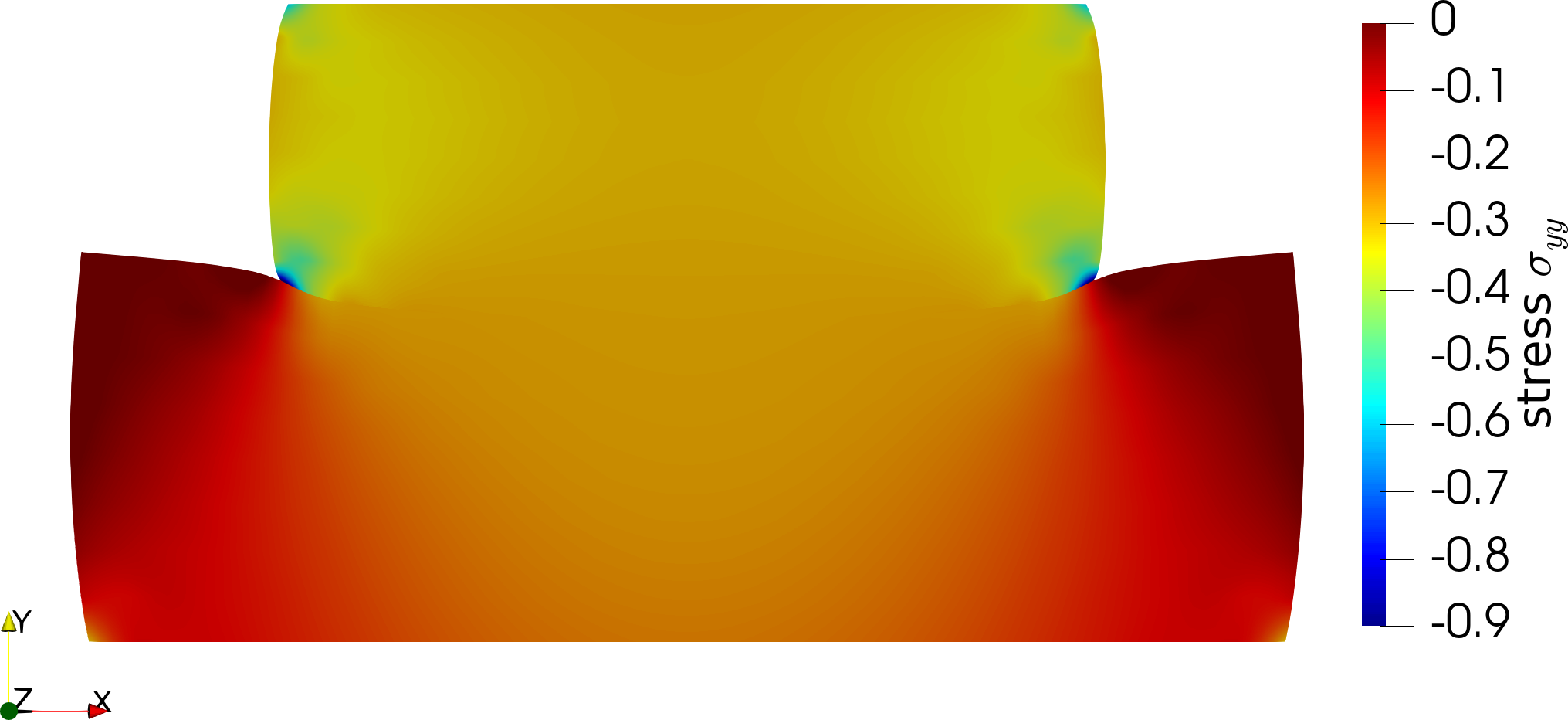}
\caption{CCS}
\end{subfigure}
\quad
\begin{subfigure}[b]{0.48\textwidth}
\includegraphics[width=\textwidth]{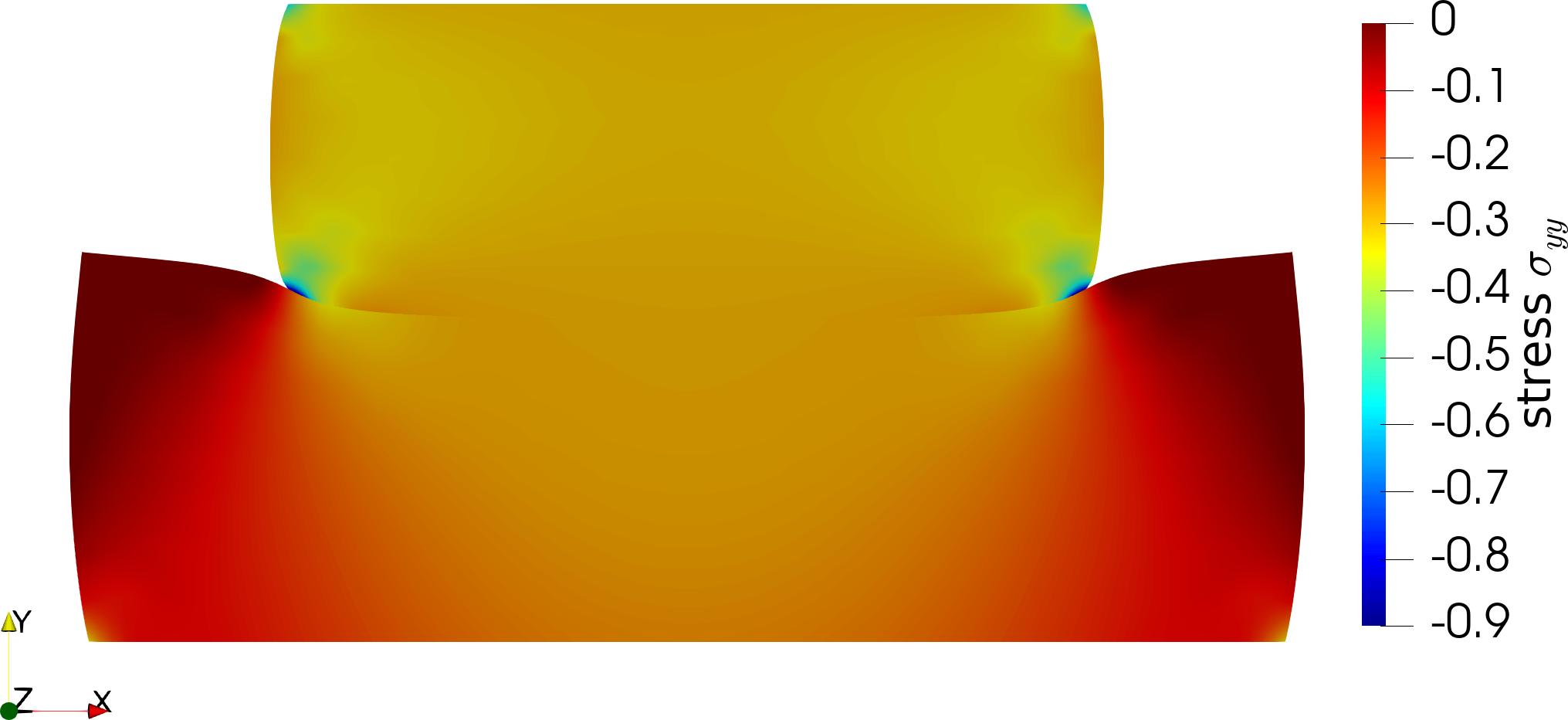}
\caption{ECCS}
\end{subfigure}\vspace{6pt}\\
\begin{subfigure}[b]{0.48\textwidth}
\includegraphics[width=\textwidth]{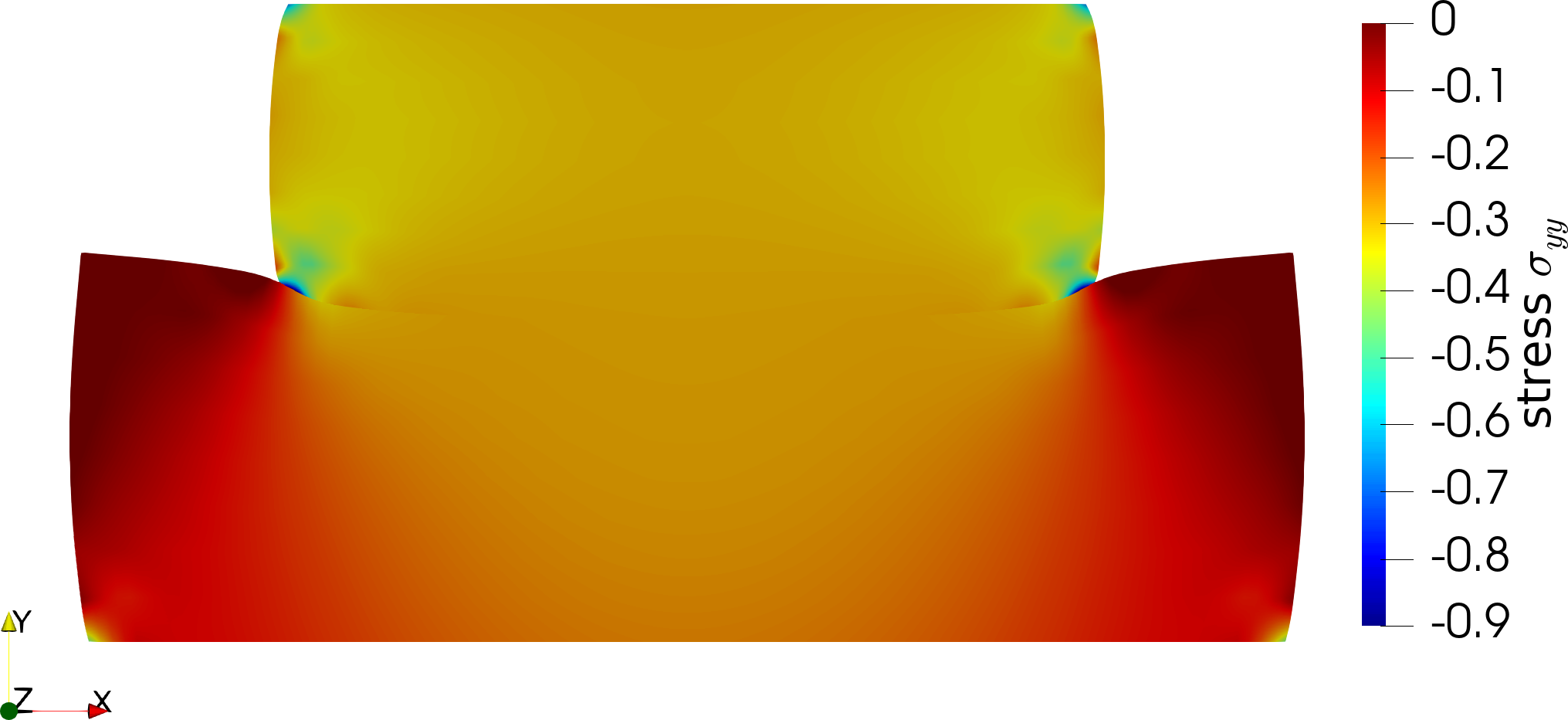}
\caption{C}
\end{subfigure}
\quad
\begin{subfigure}[b]{0.48\textwidth}
\includegraphics[width=\textwidth]{./images/solutionfields/BlockOnBlock/BlockOnBlockModcollGal_25_10_p2_Syy.png}
\caption{EC}
\end{subfigure}
\caption{Two deformable blocks: Stress $\sigma_{yy}$ for the proposed collocated contact surface approaches (CCS \& ECCS), collocation (C) and enhanced collocation (EC). Discretization of each body with $25 \times 10$ control points.}
\label{fig:StressTwoBlock2510Collocation}
\end{figure}

\begin{figure}[H]
\centering
\begin{subfigure}[b]{0.48\textwidth}
\includegraphics[width=\textwidth]{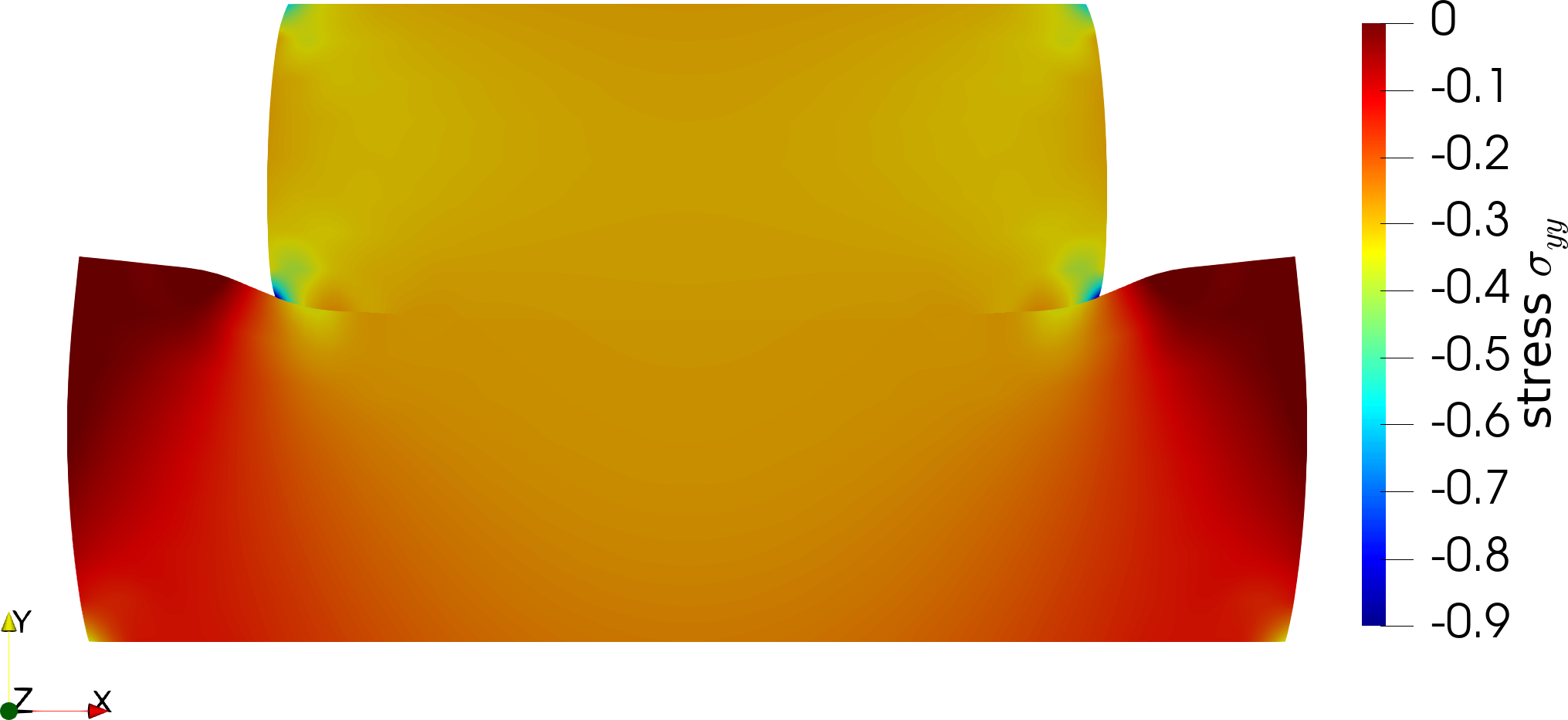}
\caption{GPTS}
\end{subfigure}
\quad
\begin{subfigure}[b]{0.48\textwidth}
\includegraphics[width=\textwidth]{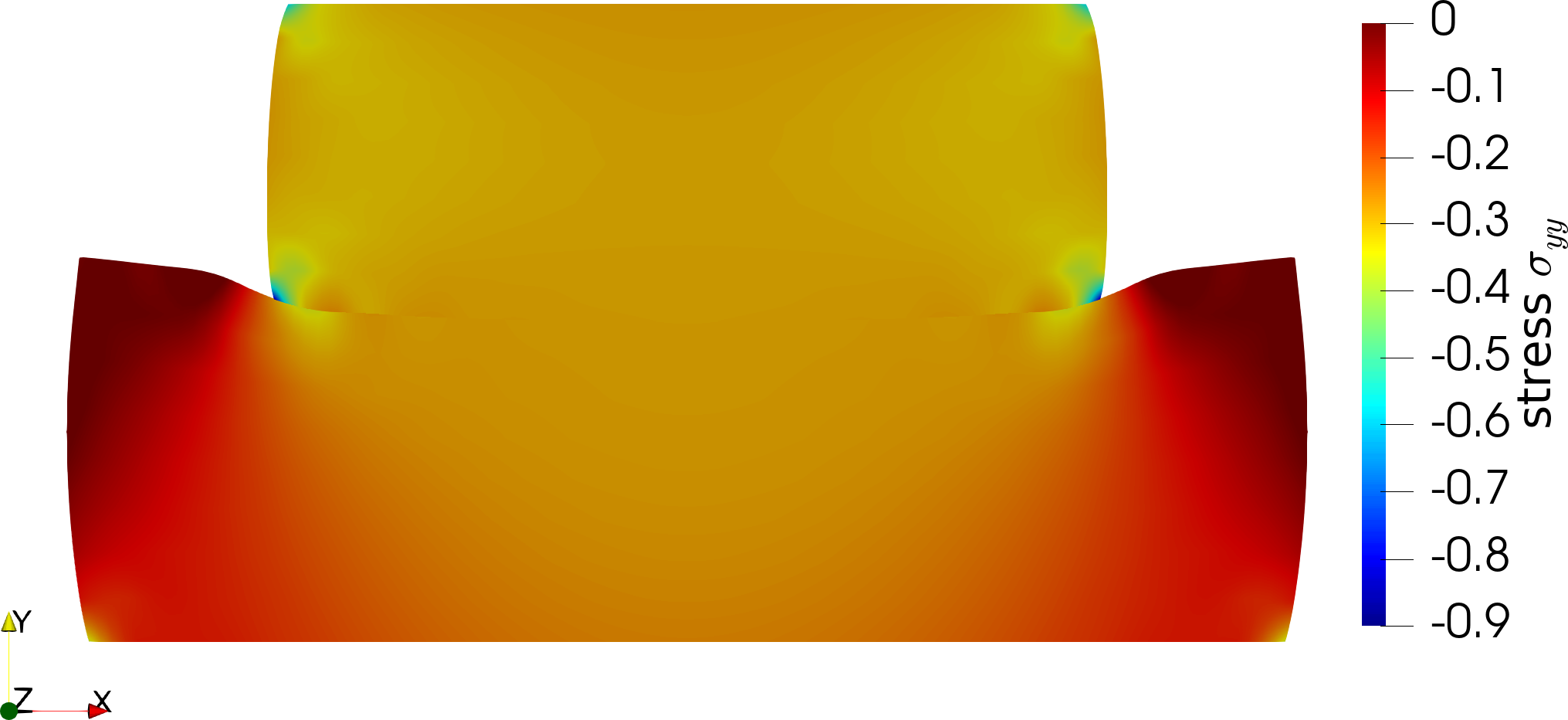}
\caption{GPTS-2hp}
\end{subfigure}\vspace{6pt}\\
\begin{subfigure}[b]{0.48\textwidth}
\includegraphics[width=\textwidth]{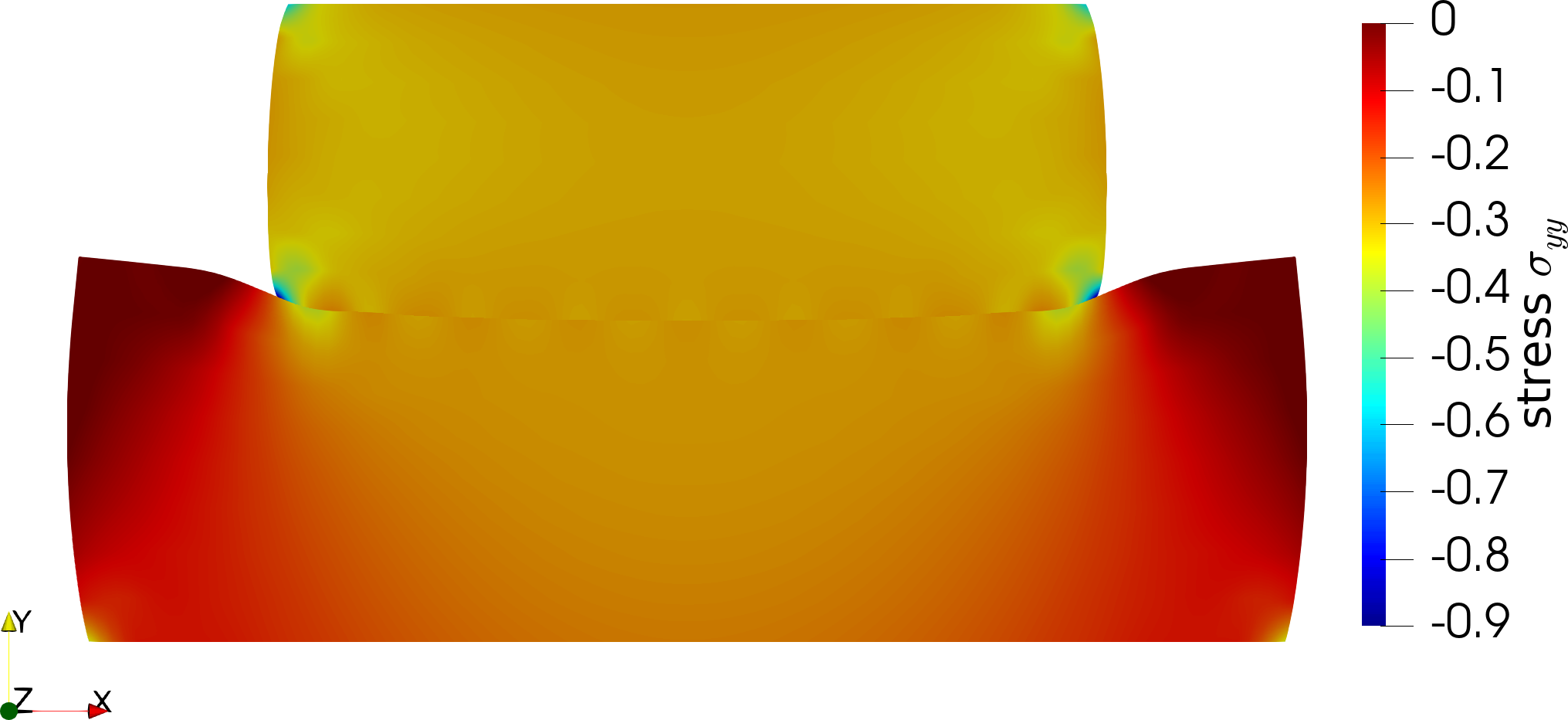}
\caption{PTS}
\end{subfigure}
\quad
\begin{subfigure}[b]{0.48\textwidth}
\includegraphics[width=\textwidth]{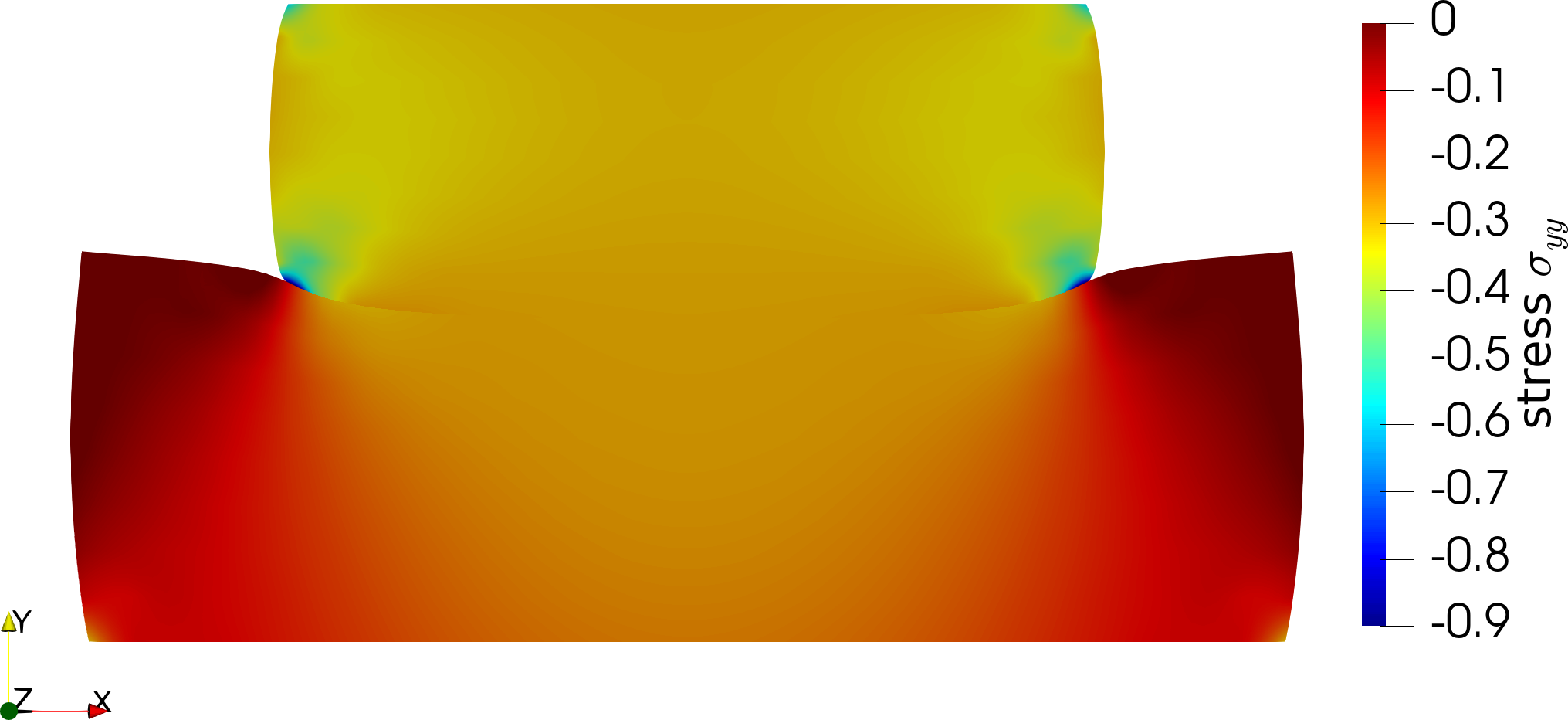}
\caption{PTS-2hp}
\end{subfigure}
\caption{Two deformable blocks: Stress $\sigma_{yy}$ for the Gauss-point-to-segment (GPTS) and Point-to-segment (PTS) approaches and the corresponding two-half-pass (2hp) formulations. Discretization of each body with $25 \times 10$ control points.}
\label{fig:StressTwoBlock2510CompMethods}
\end{figure}

In Figures \ref{fig:StressTwoBlock2510Collocation} and \ref{fig:StressTwoBlock2510CompMethods} the stress component $\sigma_{yy}$ is plotted for a finer discretization ($25\times10$ control points per body) and an aspect ratio of the elements closer to the unity than in the previous discretization. Here, also the pure collocation approach does not lead to oscillations. Interestingly, a mild checkerboard pattern appears in the contact region for the PTS approach. This effect vanishes for the corresponding two-half-pass formulation. With this exception, the obtained results are nearly identical for all methods. 

\subsection{Hertzian contact}
\label{subsec:HertzianContact}

\begin{figure}[H]
\centering
\begin{subfigure}[b]{0.38\textwidth}
\includegraphics[width=0.9\textwidth]{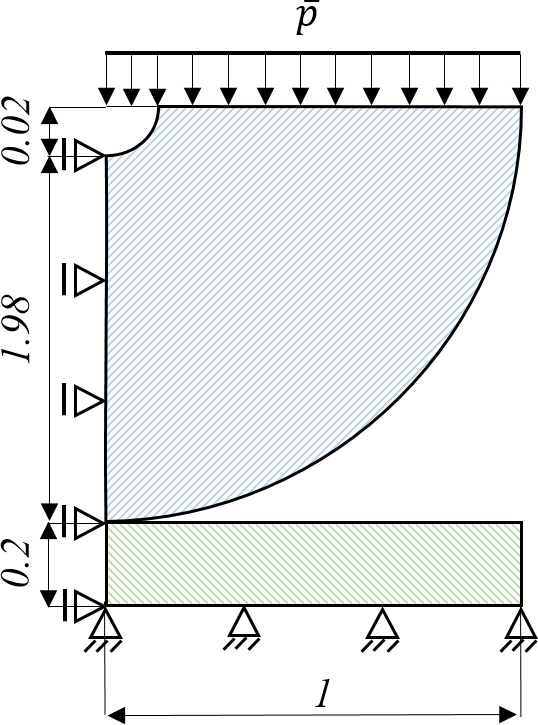}
\caption{Geometry and boundary conditions}
\end{subfigure}
\qquad
\begin{subfigure}[b]{0.42\textwidth}
\begin{mdframed}[frametitle={\footnotesize Simulation setup},userdefinedwidth=\textwidth]
\begin{footnotesize}
Number of Bézier elements: \\$n_{el} = 25\times25$ / $50\times50$\\
Polynomial degree: $p = 2-5$\\
Penalty parameter: $\epsilon_n = 1000$\\
Young’s modulus: $E = 1$\\
Poisson’s ratio: $\nu = 0.3$\\
Distributed load: $\bar{p} = 0.001$\\ 
Number of loadsteps: $n_l= 1$
\end{footnotesize}
\end{mdframed}\vspace{20pt}
\caption{Simulation parameters}
\end{subfigure}
\caption{Hertzian contact: Geometry, boundary conditions and simulation setup.}
\label{fig:GeometryHertzianContact}
\end{figure}

As a further example, the classical Hertz frictionless contact problem between a cylinder and a rigid plane is investigated. The geometry, boundary conditions and further simulation parameters are given in Figure \ref{fig:GeometryHertzianContact}. Due to the tensor product structure of the NURBS basis functions, it is necessary to model the cylinder with a small inner radius as depicted in Figure \ref{fig:GeometryHertzianContact}. The
cylinder is loaded with a vertical force P = 0.002 applied as a uniformly distributed load $\bar{p}$ on the upper surface of the cylinder and symmetry conditions are applied to the left edge. The discretization of the cylinder is refined close to the contact region by using non-uniform knot vectors such that $80\%$ of the elements are located within $10\%$ of the total length of the knot vector in both parametric directions. 

A coarse and a fine mesh are tested for four different polynomial orders to study the effect of the discretization on the results. The load is applied within one loadstep. To ensure validity of the Hertz theory, linear elasticity is assumed. 

For the considered discretizations, the dimensionless contact pressure $p/p_0$ is plotted versus the dimensionless coordinate $x/a$ in Figures \ref{fig:Hertz2525ContactPressure} and \ref{fig:Hertz5050ContactPressure}, with $a$ and $p_0$ being the half-width of the contact area and the maximum normal pressure, respectively. Although the chosen setup does not exactly correspond to the original Hertz model, the resulting error is negligible provided that the applied load is relatively small. The half-width of the contact area is calculated by the expression $a=\sqrt{\frac{4P}{\pi E'}}$ with $E'=\frac{E}{1-\nu}$ and the maximum normal pressure $p_0$ is estimated by the formula $p_0=\frac{2P}{\pi a}$. For the given setup this leads to values of $a=0.0481$ and $p_0=0.0264$. There exist procedures for the reconstruction of the contact pressures, which are able to reduce occurring oscillations in a post-processing step. Since we are mainly interested in the comparison of the different contact formulations, we reconstruct the contact pressures directly from the tractions to give an unaltered account of the performance of the tested approaches.   

The results for the coarse discretization are given in Figure \ref{fig:Hertz2525ContactPressure} along with the reference solution. Already for the coarse discretization, all obtained results are in good agreement with the reference solution. For the lowest polynomial degree, the results of the C and EC approaches show slight deviations. In the case of EC, these slight deviations vanish for the higher polynomial degrees, but they persist for the pure collocation approach. The newly proposed CCS approach is not affected by these deviations, which suggests that they may be an artefact of the incorporation of the boundary conditions and not induced by the contact formulation. 

Figure \ref{fig:Hertz5050ContactPressure} shows the results for the fine discretization, which are now nearly indistinguishable and extremely close to the analytical solution. The non-physical negative contact pressures which appear for all tested approaches close to the boundary of the contact region could be removed by a suitable post-processing scheme and are not related to a specific contact formulation. 

\begin{figure}[H]    

\begin{tikzpicture}[spy using outlines=
	{circle, magnification=1.5, connect spies}]
\begin{groupplot}[group style={group size=2 by 2, horizontal sep=1cm, vertical sep=1.5cm},width=0.45\linewidth]
    \nextgroupplot[
    title = {$\boldsymbol{p=2}$},
	axis equal image,
	every axis plot/.append style={very thick},
	grid=both,
    grid style={dashed,gray!50}, 
	legend cell align=left,
	legend style={legend pos= outer north east, font=\scriptsize},
	ylabel={$p/p_0$},
	ymin = -0.1,
	ymax = 1.1,
	xmin = 0,
	xmax = 1.4,
]
\addplot[orange,solid] table[x = x/a, y = p/p0]{\HertzAnalytical}; 
\addplot[black,dashed] table[x = x_a, y = p_p0_pu_2]{\HertzCollGalTwentyFive};
\addplot[red,densely dashed] table[x = x_a, y = p_p0_pu_2]{\HertzCollTwentyFive};
\addplot[brown,dash dot] table[x = x_a, y = p_p0_pu_2]{\HertzModCollGalTwentyFive};
\addplot[cyan,dash dot dot] table[x = x_a, y = p_p0_pu_2]{\HertzModCollTwentyFive};
\addplot[blue,dashed] table[x = x_a, y = p_p0_pu_2]{\HertzGalPePtsPlusTwentyFive};
\addplot[green,dash dot] table[x = x_a, y = p_p0_pu_2]{\HertzPureGalTwentyFive};
\coordinate (spypoint) at (axis cs:1.1,0.03);
\coordinate (magnifyglass) at (axis cs:1.85,0.2);

\nextgroupplot[
	title = {$\boldsymbol{p=3}$},
	axis equal image,
	every axis plot/.append style={very thick},
	grid=both,
    grid style={dashed,gray!50}, 
	legend cell align=left,
	legend style={legend pos= outer north east, font=\scriptsize},
	ymin = -0.1,
	ymax = 1.1,
	xmin = 0,
	xmax = 1.4,
] 
\addplot[orange,solid] table[x = x/a, y = p/p0]{\HertzAnalytical}; 
\addplot[black,dashed] table[x = x_a, y = p_p0_pu_3]{\HertzCollGalTwentyFive};
\addplot[red,densely dashed] table[x = x_a, y = p_p0_pu_3]{\HertzCollTwentyFive};
\addplot[brown,dash dot] table[x = x_a, y = p_p0_pu_3]{\HertzModCollGalTwentyFive};
\addplot[cyan,dash dot dot] table[x = x_a, y = p_p0_pu_3]{\HertzModCollTwentyFive};
\addplot[blue,dashed] table[x = x_a, y = p_p0_pu_3]{\HertzGalPePtsPlusTwentyFive};
\addplot[green,dash dot] table[x = x_a, y = p_p0_pu_3]{\HertzPureGalTwentyFive};
\coordinate (spypoint) at (axis cs:1.1,0.03);
\coordinate (magnifyglass) at (axis cs:1.85,0.2);
\legend{Ref.,CCS,C,ECCS,EC,PTS,GPTS}
 \nextgroupplot[
    title = {$\boldsymbol{p=4}$},
	axis equal image,
	every axis plot/.append style={very thick},
	grid=both,
    grid style={dashed,gray!50}, 
	legend cell align=left,
	legend style={legend pos= outer north east, font=\scriptsize},
	xlabel={$x/a$},
	ylabel={$p/p_0$},
	ymin = -0.1,
	ymax = 1.1,
	xmin = 0,
	xmax = 1.4,
]
\addplot[orange,solid] table[x = x/a, y = p/p0]{\HertzAnalytical}; 
\addplot[black,dashed] table[x = x_a, y = p_p0_pu_4]{\HertzCollGalTwentyFive};
\addplot[red,densely dashed] table[x = x_a, y = p_p0_pu_4]{\HertzCollTwentyFive};
\addplot[brown,dash dot] table[x = x_a, y = p_p0_pu_4]{\HertzModCollGalTwentyFive};
\addplot[cyan,dash dot dot] table[x = x_a, y = p_p0_pu_4]{\HertzModCollTwentyFive};
\addplot[blue,dashed] table[x = x_a, y = p_p0_pu_4]{\HertzGalPePtsPlusTwentyFive};
\addplot[green,dash dot] table[x = x_a, y = p_p0_pu_4]{\HertzPureGalTwentyFive};
\coordinate (spypoint) at (axis cs:1.1,0.03);
\coordinate (magnifyglass) at (axis cs:1.85,0.2);

\nextgroupplot[
	title = {$\boldsymbol{p=5}$},
	axis equal image,
	every axis plot/.append style={very thick},
	grid=both,
    grid style={dashed,gray!50}, 
	legend cell align=left,
	legend style={legend pos= outer north east, font=\scriptsize},
	xlabel={$x/a$},
	ymin = -0.1,
	ymax = 1.1,
	xmin = 0,
	xmax = 1.4,
] 
\addplot[orange,solid] table[x = x/a, y = p/p0]{\HertzAnalytical}; 
\addplot[black,dashed] table[x = x_a, y = p_p0_pu_5]{\HertzCollGalTwentyFive};
\addplot[red,densely dashed] table[x = x_a, y = p_p0_pu_5]{\HertzCollTwentyFive};
\addplot[brown,dash dot] table[x = x_a, y = p_p0_pu_5]{\HertzModCollGalTwentyFive};
\addplot[cyan,dash dot dot] table[x = x_a, y = p_p0_pu_5]{\HertzModCollTwentyFive};
\addplot[blue,dashed] table[x = x_a, y = p_p0_pu_5]{\HertzGalPePtsPlusTwentyFive};
\addplot[green,dash dot] table[x = x_a, y = p_p0_pu_5]{\HertzPureGalTwentyFive};
\coordinate (spypoint) at (axis cs:1.1,0.03);
\coordinate (magnifyglass) at (axis cs:1.85,0.2);
\end{groupplot}
\end{tikzpicture}
\caption{Hertzian contact: Contact pressure for discretization with $25\times25$ Bézier elements and polynomial degree $p=2,3,4,5$.}    
\label{fig:Hertz2525ContactPressure}      
\end{figure}

\begin{figure}[H]    

\begin{tikzpicture}[spy using outlines=
	{circle, magnification=1.5, connect spies}]
\begin{groupplot}[group style={group size=2 by 2, horizontal sep=1cm, vertical sep=1.5cm},width=0.45\linewidth]
    \nextgroupplot[
    title = {$\boldsymbol{p=2}$},
	axis equal image,
	every axis plot/.append style={very thick},
	grid=both,
    grid style={dashed,gray!50}, 
	legend cell align=left,
	legend style={legend pos= outer north east, font=\scriptsize},
	ylabel={$p/p_0$},
	ymin = -0.1,
	ymax = 1.1,
	xmin = 0,
	xmax = 1.4,
]
\addplot[orange,solid] table[x = x/a, y = p/p0]{\HertzAnalytical}; 
\addplot[black,dashed] table[x = x_a, y = p_p0_pu_2]{\HertzCollGalFifty};
\addplot[red,densely dashed] table[x = x_a, y = p_p0_pu_2]{\HertzCollFifty};
\addplot[brown,dash dot] table[x = x_a, y = p_p0_pu_2]{\HertzModCollGalFifty};
\addplot[cyan,dash dot dot] table[x = x_a, y = p_p0_pu_2]{\HertzModCollFifty};
\addplot[blue,dashed] table[x = x_a, y = p_p0_pu_2]{\HertzGalPePtsPlusFifty};
\addplot[green,dash dot] table[x = x_a, y = p_p0_pu_2]{\HertzPureGalFifty};

\nextgroupplot[
	title = {$\boldsymbol{p=3}$},
	axis equal image,
	every axis plot/.append style={very thick},
	grid=both,
    grid style={dashed,gray!50}, 
	legend cell align=left,
	legend style={legend pos= outer north east, font=\scriptsize},
	ymin = -0.1,
	ymax = 1.1,
	xmin = 0,
	xmax = 1.4,
] 
\addplot[orange,solid] table[x = x/a, y = p/p0]{\HertzAnalytical}; 
\addplot[black,dashed] table[x = x_a, y = p_p0_pu_3]{\HertzCollGalFifty};
\addplot[red,densely dashed] table[x = x_a, y = p_p0_pu_3]{\HertzCollFifty};
\addplot[brown,dash dot] table[x = x_a, y = p_p0_pu_3]{\HertzModCollGalFifty};
\addplot[cyan,dash dot dot] table[x = x_a, y = p_p0_pu_3]{\HertzModCollFifty};
\addplot[blue,dashed] table[x = x_a, y = p_p0_pu_3]{\HertzGalPePtsPlusFifty};
\addplot[green,dash dot] table[x = x_a, y = p_p0_pu_3]{\HertzPureGalFifty};
\coordinate (spypoint) at (axis cs:1.1,0.03);
\coordinate (magnifyglass) at (axis cs:1.85,0.2);
\legend{Ref.,CCS,C,ECCS,EC,PTS,GPTS}
 \nextgroupplot[
    title = {$\boldsymbol{p=4}$},
	axis equal image,
	every axis plot/.append style={very thick},
	grid=both,
    grid style={dashed,gray!50}, 
	legend cell align=left,
	legend style={legend pos= outer north east, font=\scriptsize},
	xlabel={$x/a$},
	ylabel={$p/p_0$},
	ymin = -0.1,
	ymax = 1.1,
	xmin = 0,
	xmax = 1.4,
]
\addplot[orange,solid] table[x = x/a, y = p/p0]{\HertzAnalytical}; 
\addplot[black,dashed] table[x = x_a, y = p_p0_pu_4]{\HertzCollGalFifty};
\addplot[red,densely dashed] table[x = x_a, y = p_p0_pu_4]{\HertzCollFifty};
\addplot[brown,dash dot] table[x = x_a, y = p_p0_pu_4]{\HertzModCollGalFifty};
\addplot[cyan,dash dot dot] table[x = x_a, y = p_p0_pu_4]{\HertzModCollFifty};
\addplot[blue,dashed] table[x = x_a, y = p_p0_pu_4]{\HertzGalPePtsPlusFifty};
\addplot[green,dash dot] table[x = x_a, y = p_p0_pu_4]{\HertzPureGalFifty};
\coordinate (spypoint) at (axis cs:1.1,0.03);
\coordinate (magnifyglass) at (axis cs:1.85,0.2);

\nextgroupplot[
	title = {$\boldsymbol{p=5}$},
	axis equal image,
	every axis plot/.append style={very thick},
	grid=both,
    grid style={dashed,gray!50}, 
	legend cell align=left,
	legend style={legend pos= outer north east, font=\scriptsize},
	xlabel={$x/a$},
	ymin = -0.1,
	ymax = 1.1,
	xmin = 0,
	xmax = 1.4,
] 
\addplot[orange,solid] table[x = x/a, y = p/p0]{\HertzAnalytical}; 
\addplot[black,dashed] table[x = x_a, y = p_p0_pu_5]{\HertzCollGalFifty};
\addplot[red,densely dashed] table[x = x_a, y = p_p0_pu_5]{\HertzCollFifty};
\addplot[brown,dash dot] table[x = x_a, y = p_p0_pu_5]{\HertzModCollGalFifty};
\addplot[cyan,dash dot dot] table[x = x_a, y = p_p0_pu_5]{\HertzModCollFifty};
\addplot[blue,dashed] table[x = x_a, y = p_p0_pu_5]{\HertzGalPePtsPlusFifty};
\addplot[green,dash dot] table[x = x_a, y = p_p0_pu_5]{\HertzPureGalFifty};
\coordinate (spypoint) at (axis cs:1.1,0.03);
\coordinate (magnifyglass) at (axis cs:1.85,0.2);
\end{groupplot}
\end{tikzpicture}
\caption{Hertzian contact: Contact stress distribution for discretization with $50\times50$ Bézier elements and polynomial degree $p=2,3,4,5$.}   
\label{fig:Hertz5050ContactPressure}            
\end{figure}

\subsection{Ironing}
\label{subsec:Ironing}

Finally, a frictionless ironing problem is studied. The setup is similar to the one described in \cite{corbett2014nurbs}. A half-cylinder is pressed into an elastic block and subsequently moved horizontally across the block. As for the Hertz problem, the half-cylinder is modeled with a small inner radius. The block is fixed on the bottom side and periodic boundary conditions are applied on the left and right sides. Both bodies are modeled as Neo-Hookean solids with the strain energy density function given in Section \ref{subsec:Kinematics}. 

In this example, in very rare cases the Newton-Raphson method was not converging, due to the residual alternating between two values in consecutive iterations (a phenomenon known as "jamming" or "zig-zagging" in the literature). In order to avoid non-converged solutions, a bisection control for the load increments was applied, so that in case of non-convergence the load increment was bisected within the corresponding loadstep.

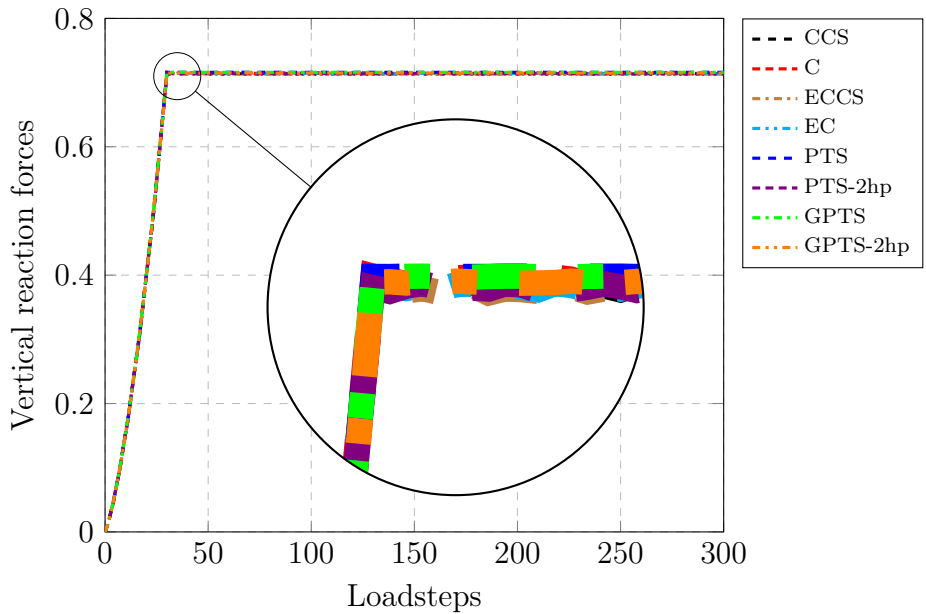
\begin{figure}[H]    

\begin{tikzpicture}[spy using outlines={circle, magnification=8, connect spies}]

\begin{axis}[
	scale=1.2,
	every axis plot/.append style={very thick},
	grid=both,
    grid style={dashed,gray!50}, 
	legend cell align=left,
	legend style={legend pos= outer north east, font=\scriptsize},
	ylabel={Vertical reaction forces},
	xlabel={Loadsteps},
	ymin = 0,
	ymax = 0.8,
	xmin = 0,
	xmax = 300,
]
\addplot[black,dashed] table[x = Loadstep, y = CollGal]{\IroningVertPTThree}; 
\addplot[red,densely dashed] table[x = Loadstep, y = Coll]{\IroningVertPThree};
\addplot[brown,dash dot] table[x = Loadstep, y = ModcollGal]{\IroningVertPThree}; 
\addplot[cyan,dash dot dot] table[x = Loadstep, y = Modcoll]{\IroningVertPThree}; 
\addplot[blue,dashed] table[x = Loadstep, y = PTS]{\IroningVertPThree}; 
\addplot[violet,densely dashed] table[x = Loadstep, y = PTSTwofield]{\IroningVertPThree}; 
\addplot[green,dash dot] table[x = Loadstep, y = GPTS]{\IroningVertPThree}; 
\addplot[orange,dash dot dot] table[x = Loadstep, y = GPTSTwofield]{\IroningVertPThree}; 
\coordinate (spypoint) at (axis cs:35,0.71);
\coordinate (magnifyglass) at (axis cs:170,0.35);
\legend{CCS, C, ECCS, EC, PTS, PTS-2hp, GPTS, GPTS-2hp}
\spy [black, size=5cm] on (spypoint)
   in node[fill=white] at (magnifyglass);
\end{axis}
\end{tikzpicture}
\caption{Ironing: Vertical reaction forces for discretization with $80\times20$ Bézier elements and polynomial degree $p=3$.}   
\label{fig:IroningVerticalReactionP3}       
\end{figure}

In Figure \ref{fig:IroningVerticalReactionP3} the vertical reaction forces are plotted. Despite the relatively coarse discretization, the curves obtained with the different algorithms are nearly indistinguishable. This test shows that the CCS approach also works well in the large deformation setting.

\begin{figure}[H]
\centering
\begin{subfigure}[b]{0.65\textwidth}
\includegraphics[width=\textwidth]{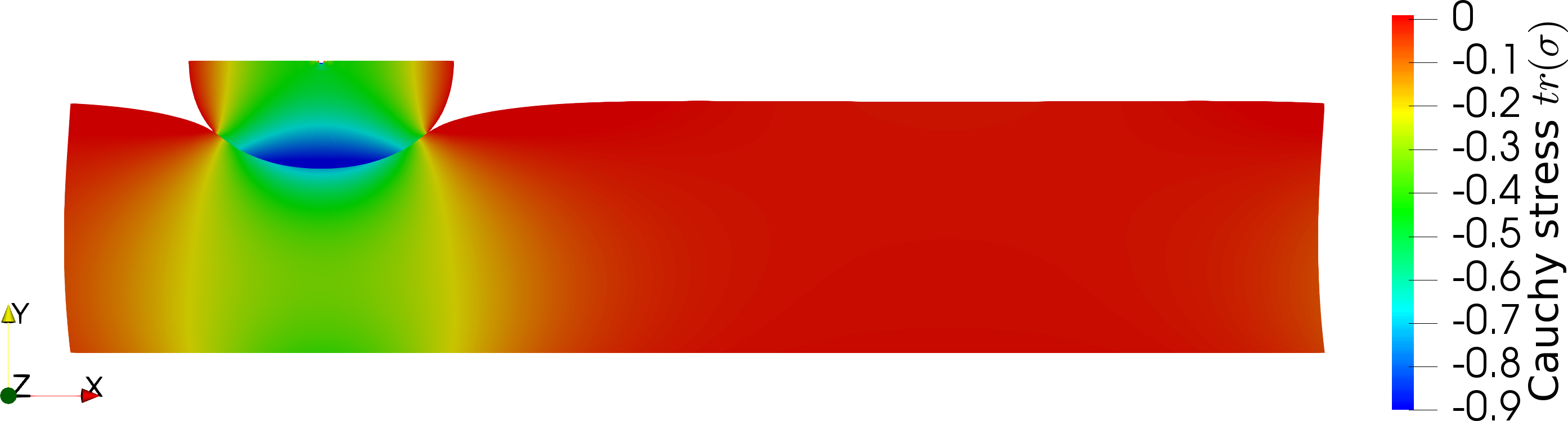}
\caption{CCS}
\end{subfigure}\vspace{6pt}\\
\begin{subfigure}[b]{0.65\textwidth}
\includegraphics[width=\textwidth]{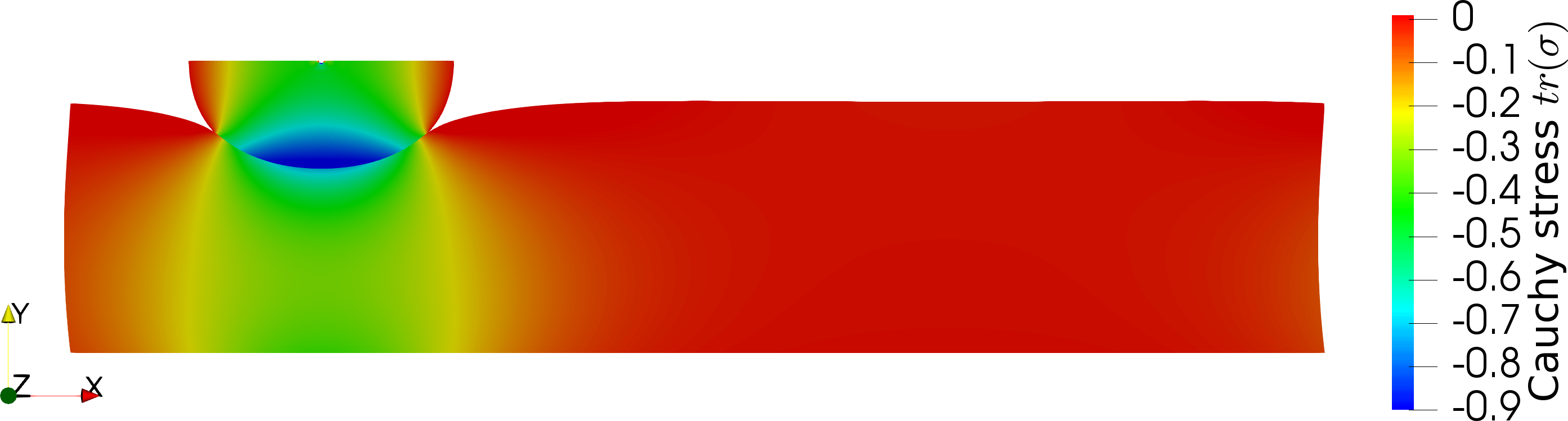}
\caption{ECCS}
\end{subfigure}\vspace{6pt}\\
\begin{subfigure}[b]{0.65\textwidth}
\includegraphics[width=\textwidth]{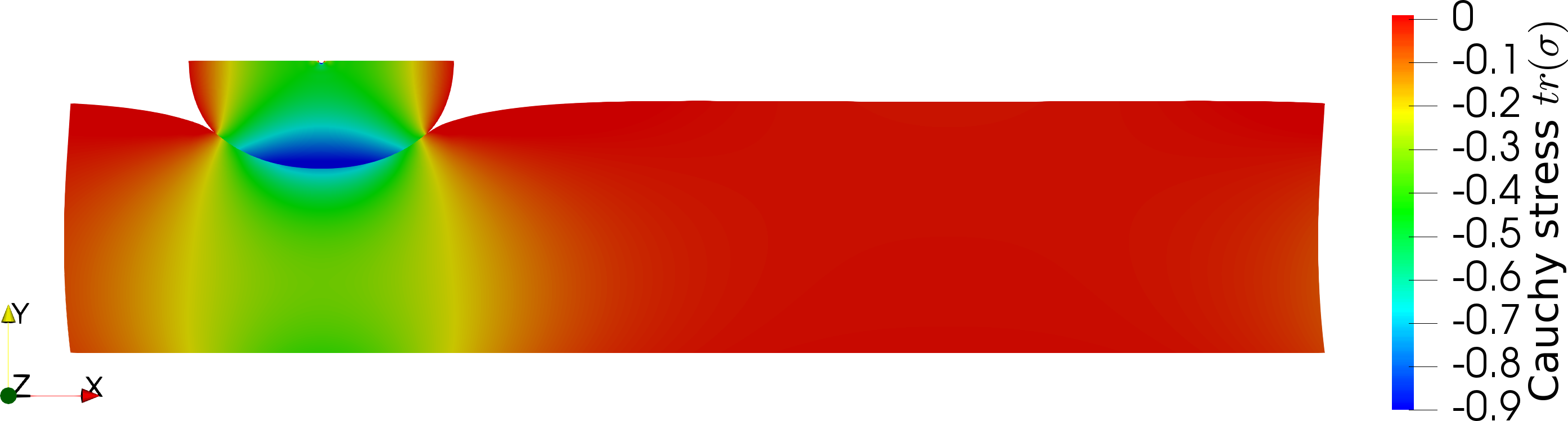}
\caption{GPTS}
\end{subfigure}\vspace{6pt}\\
\begin{subfigure}[b]{0.65\textwidth}
\includegraphics[width=\textwidth]{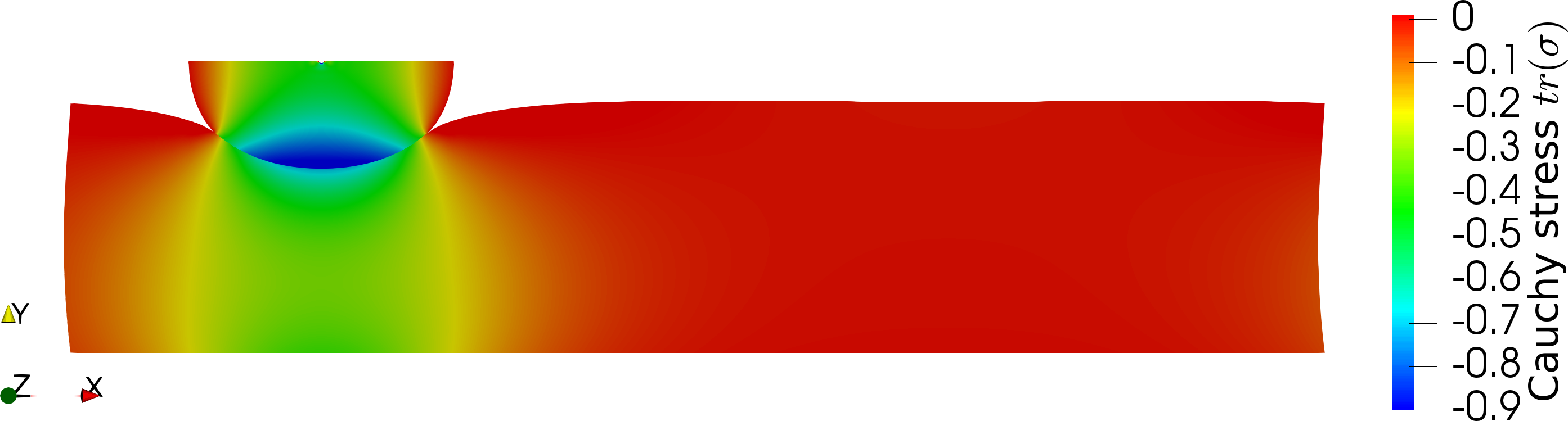}
\caption{GPTS-2hp}
\end{subfigure}

\caption{Ironing: Plots of the trace of the Cauchy stress $tr(\boldsymbol{\sigma})$ at the end of the vertical loading phase for the proposed collocated contact surface approaches (CCS \& ECCS), Gauss-point-to-segment (GPTS) and the corresponding two-half-pass (2hp) formulation. Discretization of each body with $80\times20$ Bézier elements and polynomial degree $p=3$.}
\label{fig:StressIroning8020p3Collocation}
\end{figure}

Figure \ref{fig:StressIroning8020p3Collocation} shows the trace of the Cauchy stress $tr(\boldsymbol{\sigma})$  for CCS and ECCS along with those for GPTS and the corresponding two-half-pass formulation. There are no visible differences between the plots, which further confirms the good performance of the proposed approach. 

\begin{figure}[H]
\centering
\begin{subfigure}[b]{0.64\textwidth}
\includegraphics[width=\textwidth]{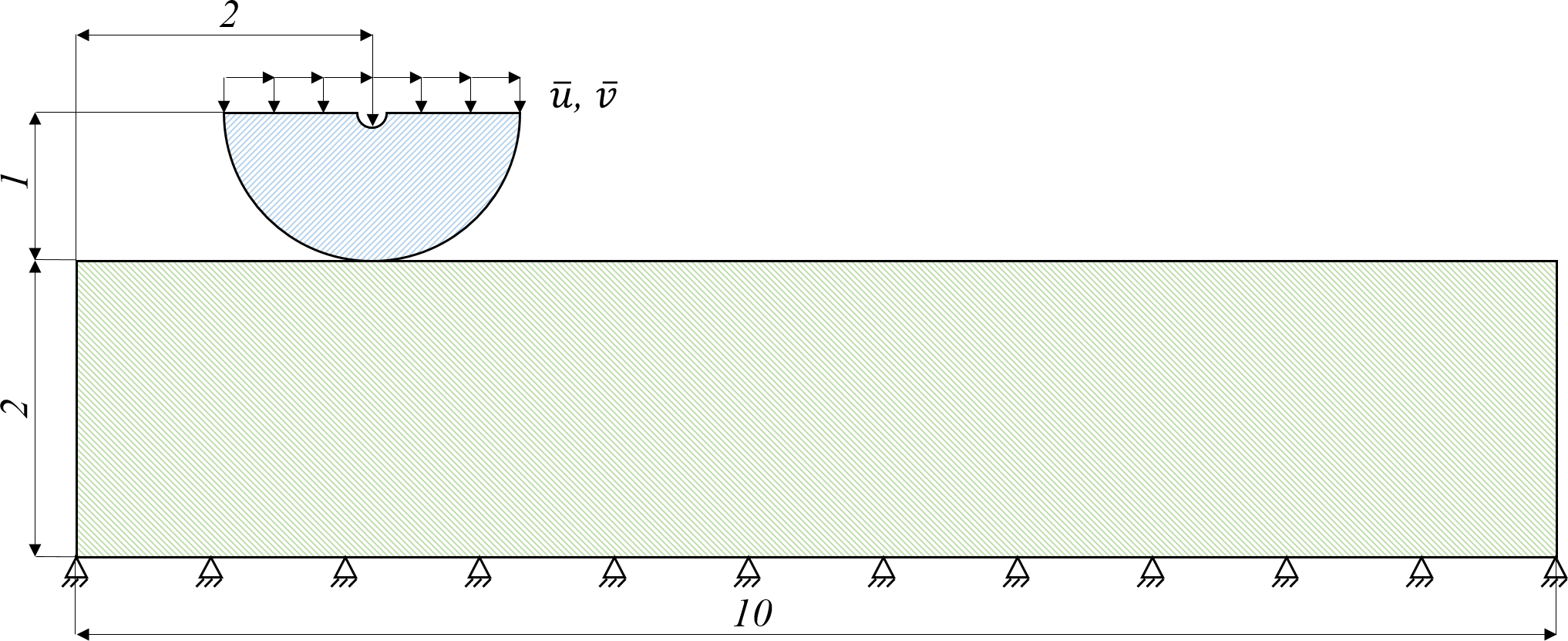}
\caption{Geometry and boundary conditions}
\end{subfigure}
\quad
\begin{subfigure}[b]{0.32\textwidth}
\begin{mdframed}[frametitle={\footnotesize Simulation setup},userdefinedwidth=\textwidth]
\begin{footnotesize}
Number of Bézier elements: \\$n_{el} = 80\times20$\\
Polynomial degree: $p = 3$\\
Penalty parameter: $\epsilon_n = 100$\\
Young’s modulus: \\ $E_{cyl.} = 3$ /  $E_{slab} = 1$ \\
Poisson’s ratio: $\nu = 0.3$\\
Number of loadsteps: \\$n_{vert.} = 30$ / $n_{horiz.} = 270$
\end{footnotesize}
\end{mdframed}
\caption{Simulation parameters}
\end{subfigure}
\caption{Ironing: Geometry, boundary conditions and simulation setup.}
\label{fig:GeometryIroning}
\end{figure}    

\section{Conclusions}
\label{sec:SummaryConclusions}

We proposed a novel hybrid discretization approach for computational contact mechanics, denoted as isogeometric Collocated Contact Surface (CCS) approach. The basic idea is to deploy the standard IGA Galerkin formulation for the bulk of deformable bodies, and to combine it with a contact formulation based on isogeometric collocation. The formulation was tested for the frictionless two-dimensionless case in both small and large deformations. Its main features, in comparison with available contact formulations, can be summarized as follows:
\begin{itemize}
\item the CCS approach is based on a simple \emph{pointwise} enforcement of the contact constraints. Unlike pointwise contact algorithms in the Galerkin framework, it passes the contact patch test to machine precision by naturally exploiting the favorable properties of isogeometric collocation;
\item compared with approaches where the discretization of both bulk and contact surfaces is based on collocation, the CCS approach does not need enhancements to remove oscillations for highly non-uniform meshes. Moreover, it enjoys the flexibility  and robustness of the Galerkin framework in the bulk discretization;
\item compared with \emph{integral} contact approaches such as Gauss-point-to-segment or mortar methods, the CCS algorithm is less expensive and easier to code, and can be added to a pre-existing isogeometric analysis code with minimal effort;
\item for frictionless contact, a drawback of the approach is the lack of symmetry of the contact contribution to the tangent stiffness matrix. However, this is no longer an issue in the more realistic situation of frictional contact, in which the tangent stiffness matrix is asymmetric in all cases.
\end{itemize}

\section{Acknowledgement}
\label{sec:Acknowledgement}
The authors gratefully acknowledge the financial support of the German Research Foundation (DFG) within the DFG Priority Program SPP 1748 “Reliable Simulation Techniques in Solid Mechanics”.

\clearpage

\bibliographystyle{elsarticle-num} 
\bibliography{./bibliography}


\end{document}